\documentclass[3p]{elsarticle} %3p,draft

\usepackage{amscd}      %AMS commutative diagrams
\usepackage{amsbsy}     %AMS bold math symbols
\usepackage{amsmath}    %AMS mathematical facilities
\usepackage{amsfonts}   %AMS Fonts
\usepackage{amsthm}     %AMS typesetting theorems
\usepackage{enumerate}  %Enumerate with redefinable labels
\usepackage{fancybox}   %Tools for using box macros.
\usepackage{geometry}   %Page layout options

\usepackage{tikz}       %Creating graphics
\usetikzlibrary{arrows,calc,decorations,shapes}

\usepackage{cancel}     %Place lines through maths formulae

\usepackage{caption}    %Customising captions in floating environments
\usepackage{fancyhdr}   %The package provides extensive facilities, both for constructing headers and footers, and for controlling their use
\usepackage{float}      %Improved interface for floating objects
\usepackage{graphics}   %Standard graphics
\usepackage{graphicx}   %Enhanced support for graphics
\usepackage{subcaption} %Support for sub-captions

\usepackage{bbm}        %Blackboard style Computer Modern fonts
\usepackage[OT1]{fontenc}%Standard package for selecting font encodings
\usepackage{bookmark}   %A new outline organization for hyperref
\usepackage{hyperref}   %Cross-referencing to produce hypertext links in the doc
\usepackage{cleveref}   %For referencing to sub-figures

\usepackage{hhline}     %Better horizontal lines in tabulars and arrays
\usepackage{longtable}  %Allow tables to flow over page boundaries
\usepackage{tabularx}   %Tabulars with adjustable-width columns

\usepackage{amsthm}
\theoremstyle{definition}
\newtheorem{theorem}{Theorem}[section]

\newtheorem{example}[theorem]{Example}
\newtheorem{remark}[theorem]{Remark}

\usepackage{eurosym}

\usepackage{algorithmicx}
\usepackage{algorithm}
\usepackage{algpseudocode}

\DeclareMathOperator*{\argmax}{arg\,max}
\numberwithin{equation}{section}	     %Equation numbering per section
\title{Valuation of electricity storage contracts using the COS method}

\begin{document}

\author[1]{Boris C. Boonstra\corref{cor1}}
\ead{borisboonstra@hotmail.com}
\author[1,2]{Cornelis W.~Oosterlee}
\ead{C.W.Oosterlee@cwi.nl}
\cortext[cor1]{Corresponding author at Delft Institute of Applied Mathematics, TU Delft, Delft, the Netherlands.}
\address[1]{CWI - National Research Institute for Mathematics and Computer Science, Amsterdam, the Netherlands}
\address[2]{Delft Institute of Applied Mathematics, Delft University of Technology, Delft, the Netherlands}

\begin{abstract}
\noindent Storage of electricity has become increasingly important, due to the gradual replacement of fossil fuels by more variable and uncertain renewable energy sources. In this paper, we provide details on how to mathematically formalize a corresponding electricity storage contract, taking into account the physical limitations of a storage facility and the operational constraints of the electricity grid. We give details of a valuation technique to price these contracts, where the electricity prices follow a structural model based on a stochastic polynomial process.  In particular, we show that the Fourier-based COS method can be used to price the contracts accurately and efficiently.
\end{abstract}

\maketitle

\section{Introduction}  \label{sec:introduction}
{\let\thefootnote\relax\footnotetext{The views expressed in this paper are the personal views of the authors and do not necessarily reflect the views or policies of their current or past employers.}}

One of the main options for the reduction of greenhouse gasses is the use of renewable energy \cite{lechtenbohmer2015decarbonising}, like wind and solar energy. The current, highly variable, output of these energy sources however creates a great challenge in maintaining a balance between demand and supply and ensuring a reliable and stable electricity network \cite{luo2015overview}.
Among all solutions for the highly variable output  \cite{julch2016comparison}, electricity storage is acknowledged as a solution with promising potential \cite{chen2009progress}. 
There are many different technologies for large-scale electricity storage systems and each technology has its own technical characteristics (see \cite{luo2015overview}, \cite{kyriakopoulos2016electrical}, \cite{poullikkas2013comparative}).
In addition, new techniques and concepts are being developed that can be used for electricity storage (e.g. Car as Power Plant \cite{lund2008integration}, \cite{ParkLee}). 

The rapid technological improvements of electricity storage are also increasingly interesting from a financial point of view, i.e., storing electricity when there is a lot of supply (and therefore a low price) and selling when the demand is high (and therefore a high price).
The business-economic consequences, profitability analyses, technological developments and applications of electricity storages have been thoroughly researched, \cite{staffell2016maximising}, \cite{lund2009role}, \cite{graves1999opportunities}, \cite{connolly2011practical}, \cite{evans2012assessment},\cite{dunn2011electrical}, \cite{chen2009progress}. 
In this paper, quantitative research is conducted into the valuation of contracts for storing electrical energy by trading on the electricity market. We formalize the contracts mathematically, taking into account the physical limitations of the electricity storage and the operational constraints of the electricity grid, while allowing the contracts to be valued efficiently by a dynamic pricing algorithm.
%Multiple profitability and business-economic analyses have been done in the literature \cite{staffell2016maximising}, \cite{lund2009role}, \cite{graves1999opportunities}, \cite{connolly2011practical}, usually for one technology in a specific application. 

Prior to the liberalization of the electricity markets, price fluctuations were minimal and often regulated. Nowadays, electricity and electricity derivatives, such as forwards and options, are traded over-the-counter (OTC) or on electricity exchanges (e.g. APX Group, Nord Pool AS, NYMEX).
Electricity prices  behave  differently  from  other commodities because electricity cannot yet be stored on a large scale (technological progress could change this) causing the prices to respond more directly to the relationship between supply and demand \cite{girish2013determinants}. This feature makes the prices exhibit unique characteristics, such as seasonality, high volatility, mean reversion and price spikes.

The early commonly used stochastic models for electricity prices were developed by Schwartz and Smith (2000) \cite{schwartz2000short} and Lucia and Schwartz (2002) \cite{lucia2002electricity}, where a two-factor model with short-term mean reverting effects was presented. Further, Barlow (2002) \cite{barlow2002diffusion} introduced a tractable structural model, based on the demand and supply, where the electricity price is a linear transformation of an underlying factor, modelled as an Ornstein-Uhlenbeck (OU) process. Many other models have been suggested for pricing in electricity markets, on which the aforementioned models often had a strong influence (cf. \cite{benth2008stochastic}, \cite{eydeland2003energy}, \cite{weron2014electricity}, \cite{carmona2014survey} for reviews on different models).

The method for modelling the electricity prices used in this paper is introduced by Filipovic, Larsson and Ware (2018) \cite{Warefilipovic2018polynomial}. They explained that with the use of a polynomial map such prices can be accurately modeled, where a polynomial process is used for the underlying factors. By means of an increasing polynomial map, the desired dynamics for the price process, e.g. creating spikes, can be mimicked, while the underlying polynomial process is typically a well-known OU process. This structural model provides a framework that shows a general and tractable relationship between the underlying factors of the polynomial process and the resulting electricity prices. 

%The unique electricity market dynamics in combination with the developing characteristics of the electricity storage and electricity grid call for accurate analysis methodologies for the potential of electricity storages.

 A dynamic pricing algorithm is based on a problem formulation with conditional expectations of the discounted value of the contract under the risk-neutral measure, the so-called risk-neutral valuation formula. 
There are several ways to compute these conditional expectations. In this paper, we focus on a Fourier-based numerical integration technique, known as the COS method, see \cite{fang2009novel}, \cite{fang2009pricing}. The main idea of the COS method is to approximate the conditional probability density function in the risk-neutral valuation formula by its Fourier cosine series expansion, where we make use of the closed-form relation between the coefficients of the Fourier cosine expansion and the characteristic function. The characteristic function of a polynomial process is generally not available, however, we use a variable transformation so that the COS method can still be employed. Furthermore, with the COS method the Greeks can be easily computed for all points in time.
Additionally, the Least Squares Monte Carlo (LSMC) method is used here to validate the results obtained with the COS method.

In Section \ref{sec:Electricity price dynamics} we introduce the stochastic price model based on polynomial stochastic processes. Subsequently, in Section \ref{sec:Electricity storage contracts} we formalize the electricity storage contracts, taking into account physical and operational constraints. Moreover, a dynamic pricing algorithm is given, allowing the contracts to be priced efficiently, and in Section \ref{sec:Valuation of the electricity storage contract} the valuation of financial derivatives on the electricity market is discussed. In Section \ref{sec:The COS method} the COS method for pricing the electricity storage contracts is presented, particularly, Subsection \ref{subsection:The COS method for electricity storage contracts} details the pricing under the polynomial process for the electricity prices. The numerical results of the electricity storage contracts are discussed in Section \ref{sec:results}, where, next to batteries, the concept Car-Park as Power Plant and the cost optimization of charging electric vehicles is analyzed. Finally, in Section \ref{sec:conclusion} we provide a conclusion reflecting on the results and the methods used.

%Introduction includes three parts; each part comprises of one paragraph:
%\begin{itemize}
%    \item Problem statement: this explains the problem and motivates why the work is essential (i.e., to provide the missing solution for a problem).
%    \item State-of-the-art: this explains what has been done to solve the problem so far, why there are still missing solutions, and what makes the work different/outstanding from current published work.
%    This paragraph first lists generic points from published work, compares it with the proposed work, and states that the proposed work can solve partly/entirely the problem.
%    \item Contributions: this first states the main contributions of the paper/articles, then lists all the detailed points of the paper’s contributions.
%    This paragraph can also be linked to the related sections in the paper, unless another paragraph is used to lists the organization of the rest of the paper.
%\end{itemize}

\section{Electricity prices and contracts} \label{sec:Electricity price dynamics}

\subsection{Electricity price dynamics}

The electricity price model, based on a so-called polynomial stochastic process, is defined as follows.
If $X_t$ follows a stochastic polynomial process, e.g. a Geometric Brownian Motion (GBM) or an Ornstein-Uhlenbeck (OU) process, then the spot price at time $t$ can be modelled in the form \cite{Warefilipovic2018polynomial}:
\begin{equation}
    S_t = \Phi (X_t) = H(X_t)^\top \mathbf{p},
    \label{Polynomial_pricing_method}
\end{equation}

where $H(X_t)$ denotes a vector of basis functions for the space of polynomials preserved by the polynomial process (e.g. $H(X_t)=(1,X_t,X_t^2,...,X_t^n)^\top$ for a one-dimensional polynomial of degree $n\in\mathbb{N}$) and the elements in vector $\mathbf{p}$ are the coefficients that characterize the polynomial map $\Phi$. The construction of these increasing polynomial maps is explained in some detail in \ref{appendix:increasing polynomial map}. Moreover, seasonality can be included by making the coefficients in vector $\mathbf{p}$ time-dependent.

Stochastic polynomial processes are relevant in many financial applications, including financial market models for interest rates, commodities and electricity. Filipovic and Larsson \cite{filipovic2016polynomial} provide a mathematical basis for polynomial processes and describe the growing range of applications in finance.

\subsection{Electricity storage contracts} \label{sec:Electricity storage contracts}
In this section, the electricity storage contracts will be defined. These are contracts where electricity can be sold/bought at fixed moments in time by trading on the electricity market. The holder of the electricity storage contract has to decide which action to take at each exercise moment. The actions that can be taken are releasing electricity from the storage, storing electricity or doing nothing. 

Compared to financial derivative contracts, as Bermudan or swing options, there are extra features to be taken into account with an electricity storage contract:
\begin{enumerate}
    \item Physical limitations of the electricity storage, e.g. capacity and endurance.
    \item Efficiency of the electricity storage.
    \item Restrictions to the amount of electricity from the grid that can be released or stored.
    \item The possibility to have negative payoff by storing electricity.
    \item A storage contract often includes a penalty function that is activated if the holder of the contract does not comply with the contract conditions.
\end{enumerate}
These additional features make the contract valuation a challenging problem. In the following two subsections, an electricity storage contract with such features is defined and the dynamic pricing algorithm for the contract valuation is presented.

\subsubsection{Details electricity storage contract}
\label{subsec:Details electricity storage contract}
In this section, a storage contract is defined, in a similar way as in \cite{boogert2008gas}, where a gas storage contract was discussed.

Consider a contract with $M$ exercise moments, where $t_0$ is the initial time and $\{t_1,...,t_M\}$ the set of exercise moments, where $0=t_0<t_1<...<t_M=T$ and the time between exercise moments is evenly distributed, $\Delta t=
\Delta t_m:=t_{m+1}-t_m$. The holder of the contract has the right to perform an action at {\em any discrete exercise time}. However, unlike most financial derivatives, the storage contract includes an extra time point, $t_{M+1}$, at which it is not possible to exercise, the so-called settlement date of the contract.

On the settlement date, the holder of the contract has to pay a penalty if the agreed amount of energy is not present in the storage. %For example, if the battery of an electric car is used for storage, there must be a certain level of energy in the battery at the end of the contract so that the car can drive. 
The penalty function at the settlement date is denoted by $q_{s}\left(t_{M+1}, S_{t_{M+1}}, {e}(t_{M+1})\right)$, where the notation ${e}(t_{m})$ represents the amount of energy left in the storage at time $t_m$, for all $m\in\{0,...,M+1\}$.

As described, the holder of the contract can take three actions: do nothing, release electricity or store electricity. These actions correspond to a change in energy level in the storage, denoted as $\Delta e(t_m)=e(t_{m+1})-e(t_m)$. If nothing is done at time $t_m$, the energy level does not change, $\Delta e(t_m)=0$, releasing electricity is defined as a negative change, $\Delta e(t_m)<0$, and storing electricity as a positive change, $\Delta e(t_m)>0$. By definition, at initial time $t_0$ the holder can not take any action and the energy change is zero, $\Delta e(t_0):=0$.

The payoff function depends on the action taken by the holder of the contract, and is defined by:
\begin{equation}
    g(t_m,S_{t_m},\Delta e(t_m))=
    \begin{cases}
        -\bar{c}(S_{t_m})\Delta e(t_m), & \Delta e(t_m)>0,\\
        0, & \Delta e(t_m)=0 ,\\
        -\bar{p}(S_{t_m})\Delta e(t_m), & \Delta e(t_m)<0,
    \end{cases}
\label{payoff function energy contract}
\end{equation}
where $\bar{c}(S_{t_m})$ and $\bar{p}(S_{t_m})$ are respectively the cost of storing and the profit of releasing electricity as a function of the electricity spot price.

Moreover, the contract includes the efficiency of the storage, e.g. an electric vehicle battery is typically between $75\%$ and $93\%$ efficient depending on the battery type \cite{ahmadian2018plug}. The efficiency of the storage will be denoted by $\eta$, which means it converts $\eta\cdot 100\%$ of the purchased amount into electricity which can be sold. Therefore, the following cost and profit functions are used:
\begin{align}
\begin{split}
        \bar{c}(S_{t_m})&=\frac{S_{t_m}}{\eta}, \\
        \bar{p}(S_{t_m})&=S_{t_m}.
\end{split}
\end{align}
These functions imply that we have to buy $(1/\eta) > 1$ units of electrical energy in order to sell $1$ unit of electrical energy.% It is  possible to add a constant to the cost and/or the profit function representing the bid-ask spread.
%It is also possible to include transaction costs and bid-ask spreads in these functions \cite{boogert2008gas}.

There are physical limitations to the capacity of the storage, for which we introduce a maximum and minimum energy level for the storage, respectively, $e^{min}$ and $e^{max}$. The energy level of the electricity storage should satisfy, for all $t_m$, $m\in\{0,...,M+1\}$:
\begin{equation}
    e^{min}\leq e(t_m) \leq e^{max}.
\end{equation}
%By allowing the capacity to be time-dependent, the physical restraints can be adjusted over time. For example, for many storages, the older a storage is, the lower the maximum capacity becomes. The temperature can also determine the minimum and maximum capacity.

Furthermore, there are operational restrictions on the minimum and maximum energy level changes at an exercise moment, respectively, $i^{min}_{op}$ and $i^{max}_{op}$. In most markets, there is also a required minimum energy to be released, denoted by $i^{min}_{market}$. So, the allowed energy level changes at an exercise moment are limited by:
\begin{equation}
    \Delta e(t_m) \in \big[i^{min}_{op},i^{min}_{market}\big] \cup \big[0,i^{max}_{op}\big], \ \ \ \ \forall m\in\{1,...,M\},
\end{equation}
where $i^{min}_{op}\leq i^{min}_{market}\leq 0\leq i^{max}_{op}$.

In addition, the endurance of the storage facility should be taken into account, e.g. charging/discharging a battery too quickly is known to reduce the battery lifetime \cite{koller2013defining}. That is why we have set an interval $[i^{min}_b,i^{max}_b]$ for energy changes in which the storage can last as long as possible. A penalty function is introduced in case the holder of the contract wants to make a change in the energy level that lies outside this interval. This penalty function for exercise moment $t_m$ depends only on the action $\Delta e(t_m)$ that is taken and is denoted as $q_b(\Delta e(t_m))$. 
\\

Summarizing the storage limitations, the allowed actions, $\Delta e(t_m)$, are limited by the capacity and the operational constraints. We define the set of allowed actions at time $t_m$, for all $m\in\{1,...,M\}$, as follows:
\begin{equation}
    \mathcal{A}(t_m,e(t_m))=\big\{\Delta e \big|  e^{min}\leq e(t_m)+ \Delta e \leq e^{max} \text{ and  } \Delta e \in \big[i^{min}_{op},i^{min}_{market}\big] \cup \big[0,i^{max}_{op}\big] \big\},
\end{equation}
and the set of allowed actions without getting a penalty $q_b(t_m, \Delta e(t_m))$ is defined by:
\begin{equation}
    \mathcal{D}(t_m,e(t_m))=\big\{\Delta e\big|  e^{min}\leq e(t_m)+ \Delta e \leq e^{max} \text{ and  } \Delta e \in \big[i^{min}_{b},i^{min}_{market}\big] \cup \big[0,i^{max}_{b}\big] \big\}.
\end{equation}
Note that the set of allowed actions where a penalty is handed out is given by $\mathcal{A}\setminus \mathcal{D}$.
\\

The value of the contract at initial time $t_0$, denoted by $v(t_0,S_{t_0})$, is given by the discounted future payoff and penalties, where the holder of the contract chooses the optimal allowed action at each exercise moment. Thus, the following pricing problem is considered:
\begin{align}
    \begin{split}
        v(t_0,S_{t_0})=&\max_{\Delta e^*}\mathbb{E}_{\mathbb{Q}} \Bigg[ \sum_{m=1}^M e^{-rt_m}\bigg(g\big(t_m,S_{t_m},\Delta e(t_m)\big) + q_b\big(\Delta e(t_m)\big)\bigg) 
        \\ &  +e^{-r t_{M+1}} q_{s}\big(t_{M+1}, S_{t_{M+1}}, {e}(t_{M+1})\big)\Bigg],
    \end{split}
\end{align}
where $\mathbb{Q}$ is the risk-neutral pricing measure, $r$ the risk-neutral interest rate and the optimal actions are given by the set $\Delta e^*=\{\Delta e^*(t_1),...,\Delta e^*(t_M)\}$.
\\
\\
In Table \ref{table: The electricity storage contract characteristics}, the characteristics of the electricity storage contract are described.
\begin{table}[H]
\centering
\begin{tabular}{l||l}
\hline
\hline
Start date                               & $t_0$                  \\
Number of exercise moments               & $M$                    \\
Time to maturity                         & $T$                    \\
Time between exercise moments            & $\Delta t=\frac{T}{M}$ \\
Settlement date                          & $t_{M+1}$              \\
Start energy level                       & $e(t_0)$              \\
Min. capacity                            & $e^{min}$              \\
Max. capacity                            & $e^{max}$              \\
Min. energy level change                & $i^{min}_{op}\leq 0$   \\
Max. energy level change                & $i^{max}_{op}\geq 0$   \\
Required min. storage in market & $i^{min}_{market}$ \\
Min. energy level change without penalty & $i^{min}_b\leq 0$        \\
Max. energy level change without penalty & $i^{max}_b\geq 0$ \\
Penalty of charging/discharging too rapidly & $q_b(\Delta e)$ \\
Penalty at settlement date               & $q_s(e)$ \\
The efficiency of the storage & $\eta$ \\
\hline
\hline
\end{tabular}
\caption{The electricity storage contract characteristics}
\label{table: The electricity storage contract characteristics}
\end{table}

\subsubsection{The contracts dynamic pricing algorithm}
This section shows how the electricity storage contract, whose details are described in Section \ref{subsec:Details electricity storage contract}, can be priced by a dynamic pricing algorithm.

First, the total electricity storage capacity is discretized into $N_e$ equally distributed energy levels, with $\delta := (e^{max}-e^{min})/N_e$ the change of energy between two consecutive energy levels. This results in the set $E:=\{e^{min}, e^{min}+\delta, e^{min}+2\delta,\cdots, e^{max}\}$ of all possible energy levels that the storage can contain. It is assumed that the action $\Delta e (t_m)$  at each exercise moment $t_m$ is a multiple of $\delta$.

Furthermore, the pricing algorithm is computed backward in time and it is not known beforehand which energy levels will be processed. Therefore, the contract values need to be computed for all possible energy levels $e\in E$. So, the notation of the contract value at each exercise moment needs to be extended by the level of energy in storage at that time. The contract value at time $t_m$, with $e(t_m)$ energy in storage and the electricity price $S_{t_m}$, is denoted by $v(t_m,S_{t_m},e(t_m))$.  

On the settlement date, $t_{M+1}$, it is not possible for the holder to change the energy level in the storage. However, if the agreed amount of energy is not present in the storage, the holder of the contract has to pay a penalty. Therefore, the value of the contract at time $t_{M+1}$ is equal to the penalty function:
\begin{equation}
    v(t_{M+1},S_{t_{M+1}}, e )=q_{s}(t_{M+1}, S_{t_{M+1}}, e ), \ \ \ \forall e\in E.
\end{equation}

At all exercise moments $t_m$, $m\in\{M,...,1\}$, before the settlement date, the holder of the contract can choose an action $\Delta e(t_m)\in \mathcal{A}$. The holder will choose the action that ultimately gives the highest value. To make this decision, continuation values are needed to determine the expected value of the contract after choosing a specific action. So, the continuation value depends on the energy level in storage after the action is taken, $e(t_{m})+\Delta e(t_m)=e(t_{m+1})$. 
%Therefore the continuation value is denoted by:
%\begin{equation}
%    c(t_m,S_{t_m},e(t_{m})+\Delta e(t_m))=c(t_m,S_{t_m},e(t_{m+1})).
%\end{equation}

Note that, at time $t_m$, the continuation value can be the same for different levels of energy in storage, by choosing certain actions. As an example, the situation with energy level $e(t_m)=2$ and action $\Delta e(t_m)=1$ will result in the same continuation value as the situation with energy level $e(t_m)=4$ and action $\Delta e(t_m)=-1$, assuming that $\Delta e(t_m)\in \mathcal{A}$.

The continuation value thus does not have to be determined for every energy level $e(t_m)$ and all corresponding allowed actions $\Delta e(t_m)\in\mathcal{A}$, but only for the possible energy levels in the set $E$. The continuation value, for all possible energy levels $e\in E$, can be computed with the risk-neutral valuation formula, i.e.,
\begin{equation}
    c(t_m,S_{t_m}, e)=e^{-r \Delta t} \mathbb{E}_{\mathbb{Q}}[v(t_{m+1},S_{t_{m+1}}, e)|\mathcal{F}_{t_m}], \ \ \ \forall e\in E,
\label{continuation value storage contract}
\end{equation}
where $\mathcal{F}_{t}=\sigma(S_s; s\leq t)$, $r$ the constant interest rate and $\mathbb{E}_{\mathbb{Q}}[\ \cdot \ ]$ the expectation under the risk-neutral measure $\mathbb{Q}$. 

Now that the continuation value has been defined, the value of the contract can be given. The holder of the contract at time $t_m$ with $e \in E$ electricity in storage and electricity price $S_{t_m}$, chooses the action $\Delta e \in\mathcal{A}(t_m,e) $ that gives the highest value, taking into account the penalty function. This results in the following contract value at time $t_m$:
\begin{align}
\begin{split}
    v(t_m,S_{t_m},e)=&\max_{\Delta e \in \mathcal{A}(t_m,e)} \Big\{  g\big(t_m,S_{t_m},\Delta e \big) +  c\big(t_m,S_{t_m}, e +\Delta e\big) + q_b\big(\Delta e \big) \Big\}, \ \forall e \in E.
\end{split}
\label{contract value formula m in M,...,1}
\end{align}

The value of the contract at initial time $t_0$ is equal to the continuation value, because no actions can be taken at this time:
\begin{equation}
     v(t_0,S_{t_0},e(t_{0}))= c(t_0,S_{t_0}, e(t_0))=e^{-r \Delta t} \mathbb{E}_{\mathbb{Q}}[v(t_{1},S_{t_{1}}, e(t_0))|\mathcal{F}_{t_0}].
\end{equation}

We now have all the building blocks to determine the contract value, backward in time. This is summarized in the following dynamic pricing algorithm:
\begin{align}
\begin{split}
    \begin{cases}
    v(t_{M+1},S_{t_{M+1}},e)&=q_s(t_{M+1},S_{t_{M+1}},e), \ \ \ \ \ \ \ \ \ \ \ \ \ \ \ \ \ \   \forall e\in E, \\
    c(t_m,S_{{t_m}}, e)&=e^{-r \Delta t} \mathbb{E}_{\mathbb{Q}}[v(t_{m+1},S_{t_{m+1}}, e)|\mathcal{F}_{t_m}]  , \  \forall e\in E \text{ and } m\in \{M, ...., 0\},\\
    v(t_m,S_{t_m},e )&=\max_{\Delta e \in \mathcal{A}} \Big\{  g\big(t_m,S_{t_m},\Delta e \big) \\
    & \ \ \ \ \ \ \ \ \ \ \ \ \ \ \ \ \  + c\big(t_m,S_{t_m}, e +\Delta e \big)
    \\ & \ \ \ \ \ \ \ \ \ \ \ \ \ \ \ \ \ + q_b\big(\Delta e \big) \Big\}, \ \ \ \ \ \ \ \ \ \ \ \ \ \  \ \forall e\in E \text{ and } m\in \{M, ...., 1\},  \\
    v(t_0,S_{t_0},e)&=c(t_0,S_{t_0}, e), \ \ \ \ \ \ \ \ \ \ \ \ \ \ \ \ \ \ \ \ \ \ \ \ \ \ \ \  \forall e\in E.
    \end{cases}
\end{split}
\label{backward induction alg electricity contract}
\end{align}

\section{Valuation of the electricity storage contract}\label{sec:Valuation of the electricity storage contract}
Option and contract pricing usually boils down to computing conditional expectations of the discounted values of the financial derivative under the risk-neutral measure, the so-called risk-neutral valuation formula. In financial mathematics, it is an important branch of research to price financial derivatives as quickly and accurately as possible. For calibration of a model, the speed of the computation is essential, while for pricing specific derivative contracts the accuracy and robustness appear crucial \cite{OosterleeGrzelak201911}. A comparison of various numerical approximation techniques was made in \cite{von2015benchop}, where Fourier-based integration techniques performed among the best for many different, but rather basic, options.

In the next section, we discuss a numerical Fourier-based valuation technique, known as the COS method, to price the electricity storage contract, where the electricity prices are defined by the structural model based on the polynomial stochastic processes.
In addition to the option value, there is also relevant information in the option Greeks, i.e., the hedge parameters or risk sensitivities.

\subsection{Risk-neutral measure}
In risk management, as well as in portfolio management, the probability space $(\Omega,\mathcal{F},\mathbb{P})$ is typically considered, where $\Omega$ denotes the sample space of all scenarios, $\mathcal{F}$ the $\sigma$-algebra on $\Omega$ and $\mathbb{P}$ the probability measure that assigns a probability to every event $\omega \in \Omega$. The probability measure $\mathbb{P}$ is called the real-world measure.

However, when pricing financial derivatives, the risk-neutral measure $\mathbb{Q}$ (also called martingale measure) is used. Under the risk-neutral measure $\mathbb{Q}$ the discounted prices of assets are martingales, which is not assured under the real-world measure $\mathbb{P}$, i.e. $\forall t\leq T$:
$$\mathbb{E}_\mathbb{Q}[e^{-r(T-t)}S_T|\mathcal{F}_t]=S_t,$$
where $\mathcal{F}_t \subset \mathcal{F}$ is the natural filtration on $(\Omega,\mathcal{F})$ and $S_t$ the price of an asset at time $t$.

The fundamental theorems of asset pricing (FTAPs) give conditions for a market to be free of arbitrage and complete.
The first FTAP states that a market is arbitrage free if and only if there exists a risk-neutral probability measure equivalent\footnote{Given a space $(\Omega, \mathcal{F})$, two measures are equivalent on $(\Omega, \mathcal{F})$ if they agree on which sets in $\mathcal{F}$ have probability zero.} to the real-world measure. The second FTAP states that an arbitrage-free market is complete, which means that every risky derivative can be hedged, if and only if there exists a unique risk-neutral measure that is equivalent to the real-world measure.

\subsubsection{Risk-neutral measure for the electricity market}
As described in the second FTAP, in complete markets the risk-neutral pricing measure is unique, ensuring only one arbitrage-free price of the option/storage contract. Furthermore, in a complete market, risk can be removed to a large extent by the delta hedging trading strategy. In contrast to a complete market, an incomplete market has many different risk-neutral measures and the property to hedge the risk is not existent. 

The electricity market is a typical example of an incomplete market, due to its characteristics, and therefore the risk-neutral measure is not unique. In the literature, there are different ways to deal with this incompleteness. It is possible to define a risk-neutral probability measure, $\mathbb{Q}$, by means of the Esscher transform. Another commonly used approach is to assume that the real-world measure is already risk-neutral, and then directly perform the pricing. This latter approach calibrates the model through implied parameters from a liquid market, however the electricity market is illiquid.

In this paper, we adopt another common approach (see e.g. \cite{lucia2002electricity}, \cite{boogert2008gas}, \cite{cartea2005pricing}), it is assumed there exists a risk premium to compensate for risk. This risk premium, defined by $\lambda \sigma(t,X_t)$, is subtracted from the real drift of the electricity price process, where $\lambda$ is the market price per unit risk linked to the state variable $X_t$ and $\sigma(t,X_t)$ the volatility parameter of the process. This risk premium, calibrated from observed market data, determines the choice of one specific risk-neutral measure. 

\section{The COS method}\label{sec:The COS method}
The risk-neutral valuation formula can be written in integral form, where $v$ denotes the option value, $x$ the state variable at time $t$ and $y$ at time $T$:
\begin{equation}
    v(t,x)=e^{-r\Delta t}  \mathbb{E}_{\mathbb{Q}}[v(T,y)|\mathcal{F}_{t}]=e^{-r\Delta t} \int_{-\infty}^{\infty} v(T,y) f(y|T,t,x) dy.
\label{Risk neutral valuation formula COS integral form}
\end{equation}

The COS method, \cite{fang2009novel}\cite{fang2009pricing}, can be applied to (underlying) processes for which the characteristic function is available. For processes with affine dynamics, the characteristic function can be derived by solving the Ricatti differential equations \cite{duffie2000transform}. Affinity is not invariant under polynomial transformations \cite{filipovic2019polynomial} and therefore the characteristic function of the model based on polynomial processes, $S_t=\Phi(X_t)$, generally does not exist. However, by using a transformation, the state variables can be conveniently chosen so that the characteristic function of the underlying process, $X_t$, can be employed.
%However, by using a transformation, the state variables, $X_t$ can be conveniently employed, so that the characteristic function of the underlying process can be used.

The conditional probability density function in equation (\ref{Risk neutral valuation formula COS integral form}) is approximated by a truncated Fourier cosine expansion, which uses the characteristic function to recover the Fourier coefficients, as follows \cite{fang2009novel}:
\begin{equation}
    f(y|T,t,x)\approx \frac{2}{b-a} \sum_{k=0}^{N-1} {}^{'}  \text{Re}\left\{ {\phi}\left(\frac{k\pi}{b-a}\bigg|\Delta t,x \right)  \cdot e^{-ik\pi \frac{a}{b-a} } \right\}  \cos\left(k\pi \frac{y-a}{b-a}\right),
    \label{approximation density fourier cosine}
\end{equation}
where $\Delta t=T-t$, $\phi(u|\Delta t,x)$ is the characteristic function of $f(y|T,t,x)$, $a$ and $b$ the truncated integration interval, $\text{Re}\{\cdot \}$ the real part of the input argument and $\sum '$ implies that the first term of the summation is multiplied by $\frac{1}{2}$. 

By replacing the conditional density function $f(y|T,t,x)$ in equation (\ref{Risk neutral valuation formula COS integral form}) by its Fourier cosine expansion approximation (\ref{approximation density fourier cosine}) and interchanging the integral and the summation, the following formula is obtained, the so-called COS formula:
\begin{equation}
    v(t,x)\approx e^{-r\Delta t} \sum_{k=0}^{N-1} {}^{'} \text{Re}\left\{ {\phi}\left( \frac{k\pi}{b-a}\bigg|\Delta t,x \right) e^{ik\pi \frac{-a}{b-a} } \right\} V_k,
    \label{COS formula European COS}
\end{equation}
where coefficients $V_k$ are defined by:
\begin{equation}
    V_k=\frac{2}{b-a}  \int_{a}^{b} v(T,y)  \cos\left(k\pi \frac{y-a}{b-a}\right) dy,
    \label{coef V_k european COS}
\end{equation}

where the state variables $x$ and $y$ can be any function of respectively the asset prices $S_{t}$ and $S_T$, e.g., in our case, $x=\Phi^{-1}(S_{t})=X_{t}$ and $y=\Phi^{-1}(S_{T})=X_{T}$. A function of the asset price that is invertible fits well with the use of the COS method, especially for the calibration of the price process parameters. A closed-form solution of the coefficients $V_k$ is available for various payoff functions and several choices of the state variables $x$ and $y$. 

The COS method will form the basis to value the electricity storage contracts.
For a large set of processes used in asset pricing models, including process $X_t$ with available characteristic function used in this paper, the characteristic function can be written in the following form:
\begin{equation}
    \phi(u|\Delta t,x)=e^{iu\beta x}\phi(u| \Delta t),
    \label{form characteristic function with beta}
\end{equation}
where $\phi(u|\Delta t)$ does not depend on $x$. For processes with independent and stationary increments, e.g. exponential L\'evy processes, it holds that $\beta=1$. For the OU process it follows that $\beta=e^{-\kappa \Delta t}$. 

%The interval of integration, $[a,b]$, for the COS method is essential for the accuracy of valuation. If an interval is set too small, a significant integration-range truncation error occurs, whereas if an interval is chosen too large, a large value for $ N $ should be selected to remain accurate, which increases the computation time.
The integration range $[a,b]$ is defined as proposed in \cite{OosterleeGrzelak201911}:
\begin{equation}
    [a,b]:=\left[\kappa_1-\bar{L}\sqrt{\kappa_2+\sqrt{\kappa_4}}, \ \kappa_1+\bar{L}\sqrt{\kappa_2+\sqrt{\kappa_4}}\right],
\end{equation}
where $\kappa_n$ is the $n^{th}$-order cumulant of the $X_t$ process and $\bar{L}$ depending on the user-defined tolerance level, typically $\bar{L}\in[6,12]$.

\subsection{The COS method for electricity storage contracts} \label{subsection:The COS method for electricity storage contracts}
In this section, the COS method for electricity storage contracts is discussed. The main idea is to approximate the continuation values, described in the dynamic pricing algorithm (\ref{backward induction alg electricity contract}), with the COS formula (\ref{COS formula European COS}) for each allowed energy level $e\in E$. 
However, the coefficients $ V_k $ also depend on the energy level $e\in E$ in storage. It follows that the COS formula for approximating the continuation value with $e \in E$ energy in storage is defined, for $m\in \{M+1,...,1\}$, by:
\begin{equation}
    c(t_{m-1},x,e)\approx \hat{c}(t_{m-1},x,e) := e^{-r\Delta t} \sum_{k=0}^{N-1} {}^{'} \text{Re}\left\{ {\phi}\left( \frac{k\pi}{b-a}\bigg|\Delta t,x \right) e^{ik\pi \frac{-a}{b-a} } \right\} V_k(t_{m},e),
    \label{COS formula for electricity storage contract}
\end{equation}
where the coefficients $V_k(t_m,e)$ are defined by:
\begin{equation}
    V_k(t_m,e)=\frac{2}{b-a}  \int_{a}^{b} v(t_m,y,e)  \cos\left(k\pi \frac{y-a}{b-a}\right) dy, \ \ \ \ \forall e\in E.
    \label{coef V_k COS electricity storage contract}
\end{equation}

The value of the electricity contract with $ e (t_0) $ electricity can be approximated if the coefficients $ V_k (t_1, e (t_0)) $ have been recovered, due to the fact that the initial value of the contract is equal to the continuation value at time $t_0$.

\subsubsection{The coefficients $V_k(t_m,e)$ for electricity storage contracts}\label{subsec: recovery V_k contract}
In this section, the coefficients $V_k(t_m,e)$, $\forall m\in\{M+1,...,1\}$ and $e\in E$, defined by formula (\ref{coef V_k COS electricity storage contract}), are recovered by means of the dynamic pricing algorithm (\ref{backward induction alg electricity contract}).

First, the coefficients are recovered at the settlement date $t_{M+1}$, where the value of the contract equals the penalty function, $v(t_{M+1},y,e) =q_s(t_{M+1}, y,e)$. Therefore, the coefficients $V_k(t_{M+1},e)$ are obtained for each energy level as follows:
\begin{equation}
    V_k(t_{M+1},e)=\frac{2}{b-a}  \int_{a}^{b} q_s(t_{M+1}, y,e)  \cos\left(k\pi \frac{y-a}{b-a}\right) dy, \ \ \ \ \forall e\in E.
    \label{coef V_k time t_M+1 COS electricity storage contract}
\end{equation}

At moments $t_m$, $m\in\{M,...,1\}$, with energy level $e\in E$, the holder of the contract chooses an allowed action $\Delta e \in \mathcal{A}(t_m,e)$ which ensures the maximum ultimate value. This decision results in the contract value $v(t_m,y,e)$ defined in (\ref{contract value formula m in M,...,1}). By substitution of this value in (\ref{coef V_k COS electricity storage contract}), the coefficients at time $t_m$, $m\in\{M,...,1\}$, with $e\in E$ energy in storage, are obtained by:
\begin{align}
\begin{split}
    V_k(t_m,e)=& \frac{2}{b-a}  \int_{a}^{b} \max_{\Delta e \in \mathcal{A}(t_m,e)} \Big\{  g\big(t_m,y,\Delta e \big) \\ &
     + c\big(t_m,y, e +\Delta e \big) + q_b\big(\Delta e \big) \Big\} \cos\left(k\pi \frac{y-a}{b-a}\right) dy, \ \ \ \ \forall e\in E.
\end{split}
\label{coef V_k time t_m m=M,...,1 COS electricity storage contract}
\end{align}

The Fourier coefficients for Bermudan-style options are obtained by the integral of a maximum of two functions \cite{fang2009pricing}. Such an integral is divided into two parts by means of the early-exercise point. When determining the coefficients for the electricity contract, $V_k(t_m,e)$ (\ref{coef V_k time t_m m=M,...,1 COS electricity storage contract}), a maximum of $Dim\big(\mathcal{A}(t_m,e)\big)$ functions is taken. 

For this, the integration range $[a,b]$ is split into intervals, $[a,x_1], [x_1,x_2],[x_3,x_4],...,[x_n,b]$, so that the maximum is always taken. This split is based on the so-called optimal actions, $\Delta e^*_i \in \mathcal{A}(t_m,e)$, for which the contract value $v(t_m,y,e)$ results in the maximum on interval $[x_i,x_{i+1}]$, for $i\in \{0,...,n\}$.

%To take the integral of the maximum in formula (\ref{coef V_k time t_m m=M,...,1 COS electricity storage contract}), the integration range $[a,b]$ is split into intervals,  $[a,x_1], [x_1,x_2],[x_2,x_3],...,[x_{n},b]$, where $\Delta e^*_i \in \mathcal{A}(t_m,e)$ is the action for which the contract value $v(t_m,y,e)$ results in its maximum on interval $[x_i,x_{i+1}]$, for $i\in \{0,...,n\}$. This action $\Delta e^*_i$, which gives the maximum contract value on the interval $[x_i,x_{i+1}]$, is called the optimal action. 

These intervals with corresponding optimal actions can be found by discretizing $[a,b]$, i.e. $[a,a+\delta,a+2\delta,...,b]$, and compare the values $v(t_m,y,e)$ for all actions $\Delta e\in\mathcal{A}(t_m,e)$ at each element in the discretization and choose the action which gives the maximum value. 

%To take the integral of the maximum in formula (\ref{coef V_k time t_m m=M,...,1 COS electricity storage contract}), the integral is split into intervals based on the actions so that the maximum is always taken. 

%To take the integral of the maximum in formula (\ref{coef V_k time t_m m=M,...,1 COS electricity storage contract}), the integration range $[a,b]$ is split into intervals, based on the so-called optimal actions $\Delta e \in \mathcal{A}(t_m,e)$, so that the maximum is always taken. This split is based on the so-called optimal actions $\Delta e^*_i \in \mathcal{A}(t_m,e)$.

Once the intervals and the corresponding optimal actions for splitting the integral are found, the coefficients $V_k(t_m,e)$ (\ref{coef V_k time t_m m=M,...,1 COS electricity storage contract}) for all $e\in E$ at moment $t_m$, $m\in\{M,...,1\}$, can be written as:
\begin{align}
    \begin{split}
        V_k(t_m,e)=&\frac{2}{b-a} \bigg[ \int_{a}^{x_1} \Big(  g\big(t_m,y,\Delta e_0^* \big)
     + c\big(t_m,y, e +\Delta e_0^* \big) + q_b\big(\Delta e_0^* \big) \Big) \cos\Big(k\pi \frac{y-a}{b-a}\Big) dy \\
     &+  \int_{x_1}^{x_2} \Big(  g\big(t_m,y,\Delta e_1^* \big)
     + c\big(t_m,y, e +\Delta e_1^* \big) + q_b\big(\Delta e_1^* \big) \Big) \cos\Big(k\pi \frac{y-a}{b-a}\Big) dy \\
     &+\  \cdots \\
     &+ \int_{x_n}^{b} \Big(  g\big(t_m,y,\Delta e_n^* \big)
     + c\big(t_m,y, e +\Delta e_n^* \big) + q_b\big(\Delta e_n^* \big) \Big) \cos\Big(k\pi \frac{y-a}{b-a}\Big) dy \bigg] \\
     =&\sum_{i=0}^n \Big[ G_k\big(x_i,x_{i+1}, \Delta e_i^* \big) +  C_k\big(x_{i},x_{i+1}, e+\Delta e_i^*\big) + Q_k\big(x_{i},x_{i+1},\Delta e_i^*\big) \Big],
    \end{split}
\end{align}
where $x_0=a$, $x_{n+1}=b$, $\Delta e_i^*\in \mathcal{A}(t_m,e)$ the optimal action on interval $[x_i,x_{i+1}]$ and the Fourier cosine series coefficients of the payoff function, the continuation value and the penalty function are respectively defined by, $\forall e\in E$ and $\Delta e\in \mathcal{A}(t_m,e)$:
\begin{align}
        G_k\big(x_i,x_{i+1}, \Delta e\big)&=\frac{2}{b-a}\int_{x_i}^{x_{i+1}} g\big(t_m,y,\Delta e \big) \cos\Big(k\pi \frac{y-a}{b-a}\Big) dy,  \label{coeffients G_k electricty} \\
        C_k\big(x_{i},x_{i+1}, t_m,e\big)&=\frac{2}{b-a}\int_{x_i}^{x_{i+1}} c\big(t_m,y, e \big)\cos\Big(k\pi \frac{y-a}{b-a}\Big) dy,  \label{coeffients C_k electricty}  \\
         Q_k\big(x_{i},x_{i+1},\Delta e\big)&=\frac{2}{b-a} \int_{x_i}^{x_{i+1}} q_b\big(\Delta e \big) \cos\Big(k\pi \frac{y-a}{b-a}\Big) dy. \label{coeffients Q_k electricty} 
\end{align}

Next, it is shown how the coefficients  $G_k(x_i,x_{i+1},\Delta e)$, $C_k\big(x_{i},x_{i+1}, e\big)$ and $Q_k(x_i,x_{i+1}, \Delta e)$ are computed, where the electricity price follows the model based on the polynomial process, $S_t=\Phi(X_t)$, as described in Section \ref{sec:Electricity price dynamics}.
\begin{remark}
It is possible to improve the computational efficiency by adding a tolerance level for a minimum integration interval of $[x_i, x_{i + 1}]$, $\forall i\in\{0,...,n\}$, i.e. if $ x_ {i + 1}-x_i<TOL$ then $[x_{i-1}, x_{i}]=[x_{i-1}, x_{i+1}]$ with optimal action $\Delta e_{i-1}^*$. By choosing a convenient tolerance level, the accuracy will be almost the same, while the computational efficiency will be significantly enhanced.
\end{remark}

\noindent \textbf{The coefficients $G_k(x_i,x_{i+1}, \Delta e)$}

\noindent The coefficients $G_k(x_i,x_{i+1},\Delta e)$ are obtained by substitution of the payoff function of the electricity contract (\ref{payoff function energy contract}) in equation (\ref{coeffients G_k electricty}), after which the integral can be calculated.

Since the characteristic function of the polynomial model $S_t=\Phi(X_t)$ is not available in general, the characteristic function of the underlying process $ X_t $ is used, which is assumed to be available. Therefore, the state variables are defined as $x=\Phi^{-1}(S_{t_{m-1}})$ and $y=\Phi^{-1}(S_{t_{m}})$. These state variables result in the following payoff function:
\begin{equation}
    g(t_m,y,\Delta e)=
    \begin{cases}
        -\frac{\Phi(y)}{\eta}\Delta e,& \Delta e>0,\\
        0, & \Delta e=0,\\
        -\Phi(y)\Delta e, & \Delta e<0.
    \end{cases}
\label{payoff function energy contract state variables}
\end{equation}
By substitution of (\ref{payoff function energy contract state variables}) in equation (\ref{coeffients G_k electricty}), the coefficients $G_k\big(x_i,x_{i+1}, \Delta e\big)$ can be calculated as follows:
\begin{equation}
    G_k\big(x_i,x_{i+1}, \Delta e \big)=
    \begin{cases}
        \frac{2}{b-a}\int_{x_i}^{x_{i+1}}\frac{ -\Phi(y)}{\eta} \Delta e \cos\Big(k\pi \frac{y-a}{b-a}\Big) dy
         ,& \Delta e>0,\\
        0 ,& \Delta e=0, \\
        \frac{2}{b-a}\int_{x_i}^{x_{i+1}} -\Phi(y) \Delta e \cos\Big(k\pi \frac{y-a}{b-a}\Big) dy, & \Delta e<0.
    \end{cases}
\label{Coef G_k for polynomial model}
\end{equation}
The coefficients $G_k\big(x_i,x_{i+1}, \Delta e \big)$ can be computed in closed-form for any finite-order polynomial map $\Phi( \cdot  )$.
\begin{example}
In this example, the coefficients $G_k\big(x_i,x_{i+1}, \Delta e\big)$ are computed where the electricity price follows a second-order polynomial model, by choosing $K=1$, $\alpha_1=0$ and $\gamma_1=\gamma$ in equation \ref{polynomial_map}, i.e.,
\begin{equation}
    S_t=\Phi(X_t)=\frac{1-\gamma}{2}X_t^2 + \gamma X_t.
\label{eq:second-order polynomial model}
\end{equation}
After substitution of the second-order polynomial model in (\ref{Coef G_k for polynomial model}),  the integral can be computed. This results in the following coefficients $G_k\big(x_i,x_{i+1}, \Delta e\big)$:
\begin{itemize}
    \item For $k=0$
    \begin{equation}
G_k\big(x_i,x_{i+1}, \Delta e\big)=
\begin{cases}
\frac{2}{b-a}\frac{\Delta e}{\eta}\Big[
-\frac{\gamma}{2}y^{2}-\frac{1-\gamma}{6}y^{3}
\Big]_{x_i}^{x_{i+1}}, &\Delta e>0,
\\
0,&\Delta e=0,
\\
\frac{2}{b-a}\Delta e\Big[
-\frac{\gamma}{2}y^{2}-\frac{1-\gamma}{6}y^{3}
\Big]_{x_i}^{x_{i+1}}, & \Delta e<0,
\end{cases}
\end{equation}
    \item For $k>0$ \\
\begin{equation}
\hspace*{-0.8cm} 
G_k\big(x_i,x_{i+1}, \Delta e\big)=
\begin{cases}
\frac{2}{b-a}\frac{\Delta e}{\eta}\left[
\begin{array}{c}
\frac{1}{2\pi^{3} k^{3}} (a-b)\bigg(\sin \Big(\frac{\pi k(a-y)}{a-b}\Big)\big(2 a^{2} (\gamma -1) - \\
 4 a b (\gamma -1)+2 b^{2} (\gamma -1)+\pi^{2} k^{2} y(y-\gamma y+2\gamma)\big) \\
 +2 \pi k(a-b)(\gamma(y-1)-y) \cos \Big(\frac{\pi k(y-a)}{a-b}\Big)\bigg)
\end{array}
\right]_{x_i}^{x_{i+1}} ,&\Delta e>0,
\\
\\
0,&\Delta e=0,
\\
\\
\frac{2}{b-a} \Delta e \left[
\begin{array}{c}
\frac{1}{2\pi^{3} k^{3}} (a-b)\bigg(\sin \Big(\frac{\pi k(a-y)}{a-b}\Big)\big(2 a^{2} (\gamma -1) - \\
 4 a b (\gamma -1)+2 b^{2} (\gamma -1)+\pi^{2} k^{2} y(y-\gamma y+2\gamma)\big) \\
 +2 \pi k(a-b)(\gamma(y-1)-y) \cos \Big(\frac{\pi k(y-a)}{a-b}\Big)\bigg)
\end{array}
\right]_{x_i}^{x_{i+1}} ,& \Delta e<0.
\end{cases}
\end{equation}
\end{itemize}
\end{example}

\noindent \textbf{The coefficients $Q_k(x_1,x_2, \Delta e)$}

\noindent The penalty function $q_b (\Delta e)$ depends only on the action $\Delta e\in \mathcal{A}$ taken and not on the electricity price. Furthermore, the penalty function will only be nonzero if an action $\Delta e \in \mathcal{A}\setminus\mathcal{D}$ is taken. Substitution of the penalty function in equation  (\ref{coeffients Q_k electricty}) results in the following Fourier cosine series coefficients of the penalty function:
\begin{itemize}
    \item For $k=0$
    \begin{equation}
        Q_k\big(x_{i},x_{i+1},\Delta e \big)=
        \begin{cases}
        0, & \Delta e \in \mathcal{D}, \\
        \frac{2}{b-a}q_b(\Delta e) (x_{i+1}-x_i), & \Delta e \in  \mathcal{A}\setminus \mathcal{D}.
        \end{cases}
    \end{equation}

    \item For $k>0$
        \begin{equation}
        Q_k\big(x_{i},x_{i+1},\Delta e \big)=
        \begin{cases}
        0, & \Delta e \in \mathcal{D}, \\
        \frac{2}{\pi k} q_b(\Delta e)\left( \sin\left(\frac{\pi k (a-x_{i+1})}{a-b}\right)-  \sin\left(\frac{\pi k (a-x_{i})}{a-b}\right)\right), & \Delta e \in  \mathcal{A}\setminus \mathcal{D}.
        \end{cases}
    \end{equation}
\end{itemize}
The Fourier coefficients of the penalty function at the settlement date (\ref{coef V_k time t_M+1 COS electricity storage contract}) are computed similarly. \\

\noindent \textbf{The coefficients $C_k(x_1,x_2, t_m,e)$}

\noindent To obtain the coefficients $C_k(x_1,x_2,t_m, e)$, the approximation of the continuation value with the COS formula, $\hat{c}(t_m,y,e)$, defined in (\ref{COS formula for electricity storage contract}), is used. Inserting $\hat{c}(t_m,y,e)$ in equation (\ref{coeffients C_k electricty}) and interchanging summation and integration gives the following approximation of coefficients $C_k(x_1,x_2, t_m,e)$, where it is assumed that the characteristic function can be written as in Equation~(\ref{form characteristic function with beta}):
%\begin{equation}
%\begin{align*}
%    \hat{C}_k(x_1,x_2, t_m) &=  \frac{2}{b-a}\int_{x_1}^{x_2} %\hat{c}(t_m,y)\cos \left(k\pi \frac{y-a}{b-a}\right) dy \\
%    &=  \frac{2}{b-a} \int_{x_1}^{x_2} e^{-r\Delta t} %\sum_{j=0}^{N-1} ^{'} \text{Re}\left\{ {\phi}\left( %\frac{k\pi}{b-a}\bigg|\Delta t,y \right) e^{ij\pi \frac{-a}{b-a} %} \right\} \cdot V_j(t_{m}) \cos \left(k\pi %\frac{y-a}{b-a}\right) dy 
%\end{align*}
%\end{equation}
%By assuming the characteristic function can be written in the %form $\phi(u;y,\Delta t)=e^{i u y \beta}\phi(y;\Delta t)$ and %interchanging summation and integration, results in the %following formula for the coefficients
\begin{equation}
    \hat{C}_k(x_1,x_2, t_m,e) =  e^{-r\Delta t} \sum_{l=0}^{N-1} {}^{'} \text{Re}\left\{ \phi\left( \frac{k\pi}{b-a}\bigg|\Delta t \right)  \hat{V}_l(t_{m+1},e)  \mathcal{M}_{k,l}(x_1,x_2) \right\},
    \label{approximation of the coefficient of the continuation value C_k}
\end{equation}
with the coefficients $ \hat{V}_l(t_{m})$, for $m=\{1,...,M\}$, given by:
\begin{equation}
    \hat{V}_l(t_{m},e)=\sum_{i=0}^n \Big[ G_l\big(x_i,x_{i+1}, \Delta e_i^* \big) +  \hat{C}_l\big(x_{i},x_{i+1}, e+\Delta e_i^*\big) + Q_l\big(x_{i},x_{i+1},\Delta e_i^*\big) \Big],
\end{equation}
where, for $m=M+1$, the coefficient $\hat{V}_l(t_{M+1},e)$ is computed as in (\ref{coef V_k time t_M+1 COS electricity storage contract}), and the coefficients $\mathcal{M}_{k,l}(x_1,x_2)$ are defined as:
\begin{equation}
    \mathcal{M}_{k,l}(x_1,x_2)=\frac{2}{b-a}\int_{x_1}^{x_2} e^{il\pi \frac{\beta y-a}{b-a}} \cos\left(k \pi \frac{y-a}{b-a}\right) dy.
\end{equation}
From basic calculus, the coefficients $\mathcal{M}_{k,l}(x_1,x_2)$ can be rewritten into two parts \cite{zhang2013efficient}:
$$\mathcal{M}_{k,l}(x_1,x_2)=-\frac{i}{\pi}\left( \mathcal{M}^s_{k,l}(x_1,x_2) + \mathcal{M}^c_{k,l}(x_1,x_2)\right),$$
where it holds that
\begin{equation}
    \mathcal{M}^c_{k,l}(x_1,x_2)=\begin{cases}
        \frac{(x_2-x_1)\pi i}{b-a}, & \text{for } k=j=0, \\
        \frac{1}{l\beta+k}\left[e^{\frac{((j\beta+k)x_2-(l+k)a)\pi i}{b-a}}- e^{\frac{((l\beta+k)x_1-(l+k)a)\pi i}{b-a}}\right], & \text{otherwise,}
    \end{cases}
    \label{M_k,l^c}
\end{equation}
\begin{equation}
    \mathcal{M}^s_{k,l}(x_1,x_2)=\begin{cases}
        \frac{(x_2-x_1)\pi i}{b-a}, & \text{for } k=j=0, \\
        \frac{1}{l\beta-k}\left[e^{\frac{((l\beta-k)x_2-(l-k)a)\pi i}{b-a}}- e^{\frac{((l\beta-k)x_1-(l-k)a)\pi i}{b-a}}\right], & \text{otherwise.}
    \end{cases}
    \label{M_k,l^s}
\end{equation}
\begin{remark}
\label{remark: FFT for bermudan option value} For characteristic functions where $\beta=1$ in Equation~(\ref{form characteristic function with beta}), an efficient fast Fourier transform (FFT) based algorithm for the computation of the Fourier coefficients of the continuation value $\hat{C}_k$ (\ref{approximation of the coefficient of the continuation value C_k}) can be used \cite{fang2009pricing}. The key to this efficient algorithm is that the matrices ${\mathcal{M}^s}=\{\mathcal{M}^s_{k,l}(x_1,x_2)\}_{k,l=0}^{N-1}$ and $\mathcal{M}^c=\{\mathcal{M}^c_{k,l}(x_1,x_2)\}_{k,l=0}^{N-1}$ have respectively a Toeplitz and Hankel structure. The algorithm to compute these coefficients with this method is described in Algorithm 2 in section 2.3 of \cite{fang2009pricing}.
\end{remark}

\subsection{The Greeks with the COS method}\label{section:The Greeks with the COS method}
The option Greeks can be calculated at almost no additional computational costs, by differentiating the COS formula. With  $v:=v(t_0,S_{t_0})$, $S:=S_{t_0}$ and $X:=X_{t_0}$, we have:
\begin{equation}
    \Delta:=\frac{\partial v}{\partial S}=\frac{\partial v}{\partial X}\frac{\partial X}{\partial S}, \ \ \ \ \ \Gamma:=\frac{\partial^2 v}{\partial S^2}=\frac{\partial^2 v}{\partial X^2}\left(\frac{\partial X}{\partial S}\right)^2 + \frac{\partial v}{\partial X}\frac{\partial^2 X}{\partial S^2}
    , \ \ \ \ \ \nu:=\frac{\partial v}{\partial \sigma},
\end{equation}
where the initial contract value of the electricity storage contract, $v(t_0,S_{t_0},e)=c(t_0,S_{t_0},e)$, is defined by the COS formula (\ref{COS formula for electricity storage contract}).
It follows that the cosine series expansions of the Greeks $\Delta$, $\Gamma$ and $\nu$ are given by:
\begin{align}
\begin{split}
    \hat{\Delta} \approx & \frac{\partial X}{\partial S} \cdot e^{-r\Delta t} \sum_{k=0}^{N-1} {}^{'} \text{Re}\left\{ {\phi}\left( \frac{k\pi}{b-a}\bigg|\Delta t,x \right) e^{ik\pi \frac{-a}{b-a} } \frac{i k \pi}{b-a}\beta \right\} V_k,  \\
    \hat{\Gamma} \approx & \left(\frac{\partial X}{\partial S}\right)^2 \cdot e^{-r\Delta t} \sum_{k=0}^{N-1} {}^{'} \text{Re}\left\{ {\phi}\left( \frac{k\pi}{b-a}\bigg|\Delta t,x \right) e^{ik\pi \frac{-a}{b-a} } \left(\frac{i k \pi}{b-a}\beta \right)^2 \right\} V_k \\
    & \ \ + \frac{\partial^2 X}{\partial S^2} \cdot e^{-r\Delta t} \sum_{k=0}^{N-1} {}^{'} \text{Re}\left\{ {\phi}\left( \frac{k\pi}{b-a}\bigg|\Delta t,x \right) e^{ik\pi \frac{-a}{b-a} } \frac{i k \pi}{b-a}\beta \right\} V_k, \\
    \hat{\nu}\approx & \  e^{-r\Delta t}  \sum_{k=0}^{N-1} {}^{'} \text{Re}\Bigg\{ \frac{\partial \phi\big( \frac{k\pi}{b-a}\big|\Delta t,x \big)}{\partial \sigma} e^{ik\pi \frac{-a}{b-a} } \Bigg\} V_k ,
\end{split}
\label{Greeks COS}
\end{align}
where the characteristic function is defined as in (\ref{form characteristic function with beta}).
This formula can also be used to determine the Greeks at exercise moments in the future, after the optimal action has been taken, making the option value equal to the continuation value, for fixed energy level and electricity price.

The Greeks (\ref{Greeks COS}) of the electricity storage contract, where the price process follows the polynomial model $S_t=\Phi(X_t)$, can be computed as we assume the inverse of the polynomial map $\Phi^{-1}(S_t)$ exists.

\section{Results and discussion} \label{sec:results}
In this section, several different electricity storage contracts are valued and the Greeks are computed by the COS method. In addition, by using the Least Squares Monte Carlo (LSMC) method \cite{longstaff2001valuing, dorr2003valuation}, the $95\%$ confidence intervals of the contract values are determined and the average energy level in storage for each contract over time is given. The LSMC method can also be applied to value the electricity storage contracts (see \ref{appendix: LSMC algorithm}).

For pricing various electricity storage contracts the electricity price is modeled here according to a second-order polynomial model, as defined in equation (\ref{eq:second-order polynomial model}), i.e.,
\begin{equation}
    S_t=\Phi(X_t)=\frac{1-\gamma}{2}X_t^2 + \gamma X_t,
    \label{S_t electricity contract}
\end{equation}
where the underlying process $ X_t $ follows an OU process (see \ref{appendix:OU process} for details):
$$dX_t= \kappa(\theta - X_t)dt + \sigma dW_t.$$
To generate an electricity price with more extreme spikes, a jump process can be added to the OU process.

The parameters for this model are chosen, so that the electricity price modeled is representative of a real electricity market with different volatilities:
\begin{equation}
\gamma=0.5, \ \ \ \ \kappa=0.3, \ \ \ \ \theta=10.1, \ \ \ \ \sigma\in \{0.3,0.6,0.9,1.2\}, \ \ \ \ X_0=10, \ \ \ \ S_0=\Phi(X_0),
\label{parameters electricity contract}
\end{equation}
and the risk-free interest rate $r=0.01$.

\begin{figure}[H]
\centering
    \begin{tabular}{c c}
        \ \ \ \  \textbf{Low volatility} & \ \ \ \ \textbf{Mid-low volatility} \\
        \includegraphics[width=5.5cm]{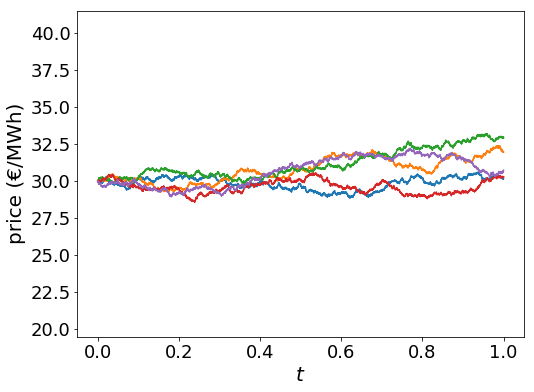} & \includegraphics[width=5.5cm]{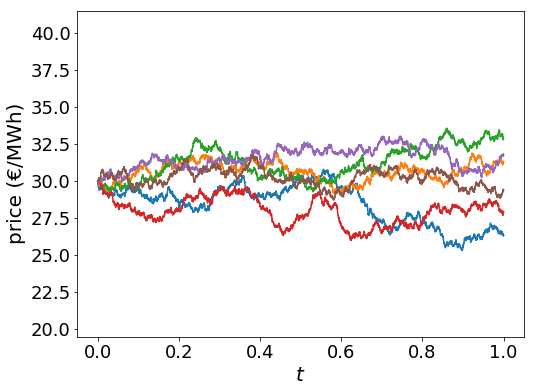} \\
        \ \ \ \  \textbf{Mid-high volatility} & \ \ \ \ \textbf{High volatility} \\
        \includegraphics[width=5.5cm]{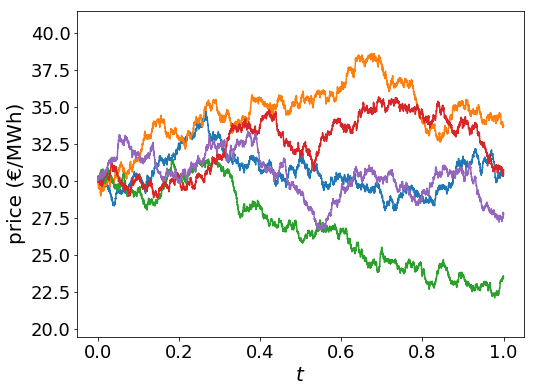} & \includegraphics[width=5.5cm]{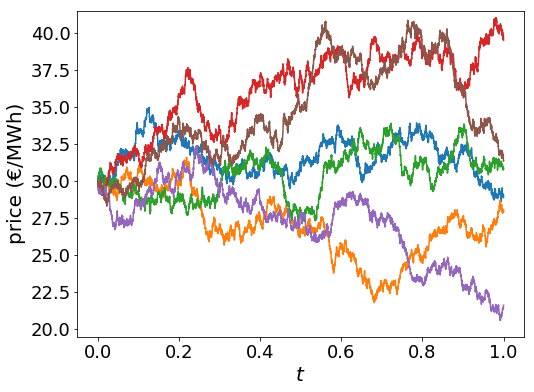} 
    \end{tabular}
\caption{Six simulations of the electricity price process simulated by the second-order polynomial model (\ref{S_t electricity contract}) with parameters defined in (\ref{parameters electricity contract}).}
\label{figure: simulation high and low volatility}
\end{figure}

The valuation of the electricity storage contract is done for $\sigma=0.3$ (low volatility),  $\sigma=0.6$ (mid-low volatility), $\sigma=0.9$ (mid-high volatility)  and $\sigma=1.2$ (high volatility). Note that for a stock market these are fairly high volatility parameters, however the electricity market is much more volatile than the stock market.
 In Figure \ref{figure: simulation high and low volatility}, six simulations of the electricity price process trajectory $ S_t $ are shown for each volatility parameter.

%\begin{figure}[H]
%\centering
%    \begin{tabular}{c c c}
%        \ \ \ \  \textbf{Low volatility} & \ \ \ \ \textbf{Mid volatility} & \ \ \ \ \textbf{High volatility} \\
%        \includegraphics[width=4.5cm]{figures/price_sigma06.png} & %\includegraphics[width=4.5cm]{figures/price_sigma09.png} &
%        \includegraphics[width=4.5cm]{figures/price_sigma12.png}
%    \end{tabular}
%\caption{Six simulations of the electricity price process simulated by the second-order polynomial model %(\ref{S_t electricity contract}) with parameters defined in (\ref{parameters electricity contract}).}
%\label{figure: simulation high and low volatility}
%\end{figure}
\subsection{The electricity storage contracts}\label{section:The electricity storage contracts}
This section is focused on four different electricity storage contracts.  There are many different technologies for large scale electricity storage systems and each technology has its own technical characteristics. Each contract is based on some of these technical characteristics. In addition, expected improvements of the characteristics and new concepts to store electricity are considered, such as higher battery efficiency, the concept of the Car-Park as Power Plant \cite{ParkLee} and cost optimization of charging electric vehicles.

Some of the electricity storage contract characteristics will be the same for each contract, these are shown in Table \ref{table: general parameters}. The other contract characteristics will differ per contract.
\begin{table}[H]
\centering
\begin{tabular}{l||l|ll}
\hline
\hline
Start date                               & $t_0$& $0$ &\\
Time to maturity                         & $T$ & $1$    &                \\
Number of exercise moments               & $M$& $50$     &               \\
Time between exercise moments            & $\Delta t$ & $1/50$  &\\
Settlement date & $t_{M+1}$ &  $T+\Delta t$ \\
Required min. release in market & $i^{min}_{market}$ & $-0.1$ &MWh \cite{ParkLee}\\
\hline
\hline
\end{tabular}
\caption{The general electricity storage parameters, used for each discussed contract.}
\label{table: general parameters}
\end{table}

\noindent \textbf{Contract 1: Standard electricity storage}\\
The most widely used electricity storage system is the rechargeable battery. The current rechargeable battery storage facilities often have capacities between 0.25-50 MWh and an output of 0.1-20 MW, depending on the battery \cite{palizban2016energy}.
Furthermore, rechargeable batteries can have an efficiency up to $\sim 95\%$ \cite{julch2016comparison}. In addition, charging and discharging a rechargeable battery too rapidly reduces the battery lifetime, so there is a penalty $q_b(\Delta e)<0$ for this. Moreover, the settlement penalty of $350$ euro is activated when the energy level in the battery is lower than the level when the contract started.

The characteristics for contract 1 are given in Table \ref{table: characteristics contract 1}.
\begin{table}[H]
\centering
\begin{tabular}{l||l | l l }
\hline
\hline
Start energy level                       & $e(t_0)$  &  $7$ & \text{MWh} \\
Min. capacity                            & $e^{min}$ & $0$ &\text{MWh}            \\
Max. capacity                            & $e^{max}$&$15 $ &\text{MWh}              \\
Min. energy level change                & $i^{min}_{op}$&$-6 $ &\text{MWh}   \\
Max. energy level change                & $i^{max}_{op}$&$6 $ &\text{MWh}   \\
Min. energy level change without penalty & $i^{min}_b$&$-4 $  &\text{MWh}        \\
Max. energy level change without penalty & $i^{max}_b$&$4 $ &\text{MWh}  \\
Efficiency of the battery & $\eta$& $ 95 $ & $\%$ \\
Penalty of charging/discharging too rapidly & $q_b(\Delta e)$ for $\Delta e\in \mathcal{A}\setminus\mathcal{D}$ & $-3$ & \EUR{} \\
Penalty at settlement date &  $q_s(e)$ for $e<e(t_0)$ & $-350$ & \EUR{} \\
\hline
\hline
\end{tabular}
\caption{The electricity storage characteristics for contract 1.}
\label{table: characteristics contract 1}
\end{table}

\noindent  \textbf{Contract 2: Highly efficient electricity storage}\\
In this contract, a highly efficient electricity storage is considered, with an efficiency of $100\%$. Batteries of $\sim 100 \%$ efficiency already exist, but due to the high manufacturing costs, they are not yet used for electricity storage \cite{chen2009progress}. The contract characteristics for contract 2 are the same as for contract 1  (Table \ref{table: characteristics contract 1}), only the efficiency of the electricity storage is increased to $\eta=100\%$.
\\

\noindent \textbf{Contract 3: Car-Park as Power Plant (CPPP)}\\
The Car-Park as Power Plant is a concept that can be seen as a solution for the highly variable output of renewable energy sources. Cars are parked for an average of $ 96\%$ of the time \cite{kempton1997electric}, therefore there is potential for using the batteries of electric vehicles for electricity storage. A single car cannot participate on the electricity market because there are minimal grid injection rules of $100$ KWh, while a full electric car has an average capacity of $80$ KWh. In addition, a single car may not be continuously available, while the presence of a large number of cars is highly predictable \cite{kempton2001vehicle}. That is why we consider multiple electric vehicles at the same time that are seen as one storage facility, e.g. a car park can be used \cite{ParkLee}.

For this contract, a car park of $150$ electric vehicles and an average of $80$ KWh capacity and $90\%$ efficiency per electric vehicle is considered.  At each exercise moment, the vehicle's battery can change $25\%$ of its energy level without getting a penalty. This results in a total capacity of $12$ MWh and an energy change without any penalty of $3$ MWh.

At the end of the contract, in our setting, the owner of the car must be able to drive, which is why there is a high penalty if there is not enough energy in the car battery.

\begin{table}[H]
\centering
\begin{tabular}{l||l | l l }
\hline
\hline
Start energy level                       & $e(t_0)$  &  $6$ & \text{MWh} \\
Min. capacity                            & $e^{min}$ & $0$ &\text{MWh}            \\
Max. capacity                            & $e^{max}$&$12 $ &\text{MWh}              \\
Min. energy level change                & $i^{min}_{op}$&$-4 $ &\text{MWh}   \\
Max. energy level change                & $i^{max}_{op}$&$4 $ &\text{MWh}   \\
Min. energy level change without penalty & $i^{min}_b$&$-3 $  &\text{MWh}        \\
Max. energy level change without penalty & $i^{max}_b$&$3 $ &\text{MWh}  \\
Efficiency of the battery & $\eta$& $ 90 $ & $\%$ \\
Penalty of charging/discharging too rapidly & $q_b(\Delta e)$ for $\Delta e\in \mathcal{A}\setminus\mathcal{D}$ & $-10$ & \EUR{} \\
Penalty at settlement date &  $q_s(e)$ for $e<e(t_0)$ & $-2000$ & \EUR{} \\
\hline
\hline
\end{tabular}
\caption{The electricity storage characteristics for contract 3.}
\label{table: characteristics contract 5}
\end{table}

\noindent \textbf{Contract 4: Cost optimization of charging electric vehicles}\\
Besides viewing the battery of electric vehicles as a storage facility, the contract can also be used to examine how the battery can be charged in a cheap way. In addition to the savings, the cost optimization eases the pressure on the electricity grid during peak power demand, because the electricity price is high then and lower during periods of low demand \cite{lunz2012influence}.

This price-based charging approach can be applied to places where multiple electric vehicles are continuously parked and need to be charged at the end of a period. % such as a company car park or an airport parking lot. 
In this contract a similar car park as in contract 3 is assumed. 
%However, now we can only charge the vehicle, instead of both charging and discharging. Although in the numerical results we also look at the price difference if both buying and selling energy is allowed, instead of just buying. 

At the end of the contract, the car batteries should be fully charged, therefore there is an increasing penalty if the battery has less than the maximum capacity on the settlement date and if the energy is lower than a certain level, defined by $e_{fix}$, there is a fixed higher penalty to pay. This penalty function, where we set $e_{fix}=6$, is defined by:    
\begin{equation}
        q_s(e)=
        \begin{cases}
        -1000 \cdot \frac{e^{max}-e}{e^{max}-e_{fix}}, & e\geq e_{fix}, \\
        -2000, & e<e_{fix}.
        \end{cases}
    \label{penalty_settlement_function_penalty_cheap_charging}
\end{equation}
The characteristics of this contract are defined in Table \ref{table: characteristics contract 4:cheap charge}. In the numerical results, we also compare this contract with a contract where the cars are unable to ``release'' energy, i.e. $i^{min}_{op}=i^{min}_b=0$, while the other characteristics remain the same.

\begin{table}[H]
\centering
\begin{tabular}{l||l | l l }
\hline
\hline
Start energy level                       & $e(t_0)$  &  $2$ & \text{MWh} \\
Min. capacity                            & $e^{min}$ & $0$ &\text{MWh}            \\
Max. capacity                            & $e^{max}$&$12 $ &\text{MWh}              \\
Min. energy level change                & $i^{min}_{op}$&$-4 $ &\text{MWh}   \\
Max. energy level change                & $i^{max}_{op}$&$4 $ &\text{MWh}   \\
Min. energy level change without penalty & $i^{min}_b$&$-3 $  &\text{MWh}        \\
Max. energy level change without penalty & $i^{max}_b$&$3 $ &\text{MWh}  \\
Efficiency of the battery & $\eta$& $ 90 $ & $\%$ \\
Penalty of charging/discharging too rapidly & $q_b(\Delta e)$ for $\Delta e\in \mathcal{A}\setminus\mathcal{D}$ & $-10$ & \EUR{} \\
Penalty at settlement date &  $q_s(e)$ & \text{see } (\ref{penalty_settlement_function_penalty_cheap_charging}) & \EUR{} \\
\hline
\hline
\end{tabular}
\caption{The electricity storage characteristics for contract 4.}
\label{table: characteristics contract 4:cheap charge}
\end{table}

\subsection{The numerical contract values}
This section shows the numerical results of the valuation of the electricity storage contracts, which are defined in Section \ref{section:The electricity storage contracts}, obtained with the COS method. Moreover, the $95\%$ confidence intervals of the contract values are given, derived with the LSMC method. The confidence interval is constructed as follows:
\begin{equation}
\text{Confidence Interval}=\left[\overline{V}-z_{\alpha/2}\left(\frac{\bar{\sigma}}{\sqrt{10}}\right) , \overline{V}+z_{\alpha/2}\left(\frac{\bar{\sigma}}{\sqrt{10}}\right) \right],
\label{Confidence interval LSMC}
\end{equation}
where $\overline{V}$ is the sample mean of ten experiments, $\bar{\sigma}$ the standard deviation and $z_{\alpha/2}$ the critical Z-value ($z_{\alpha/2}=1.96$ for a $95\%$ confidence interval). The resulting confidence intervals are computed by ten runs with the LSMC method of $25\, 000$ trajectories.

Furthermore, the average energy levels and the $95\%$ confidence intervals of the energy levels in the storage are shown. The average energy level gives an indication to the holder of a contract which energy levels must be maintained to get maximum profit. Additionally, the maximum and minimum energy levels of all trajectories used are given.

For the COS method, we use the terms $N\in\{100,150,200\}$ in the Fourier series expansion and $\bar{L}=10$ to define the integration interval.  In addition, the maximum and minimum energy levels at each exercise moment are taken over all trajectories of the ten runs of the LSMC method. 

\begin{figure}[H]
\centering
    \begin{tabular}{c c}
        \includegraphics[width=5.3cm]{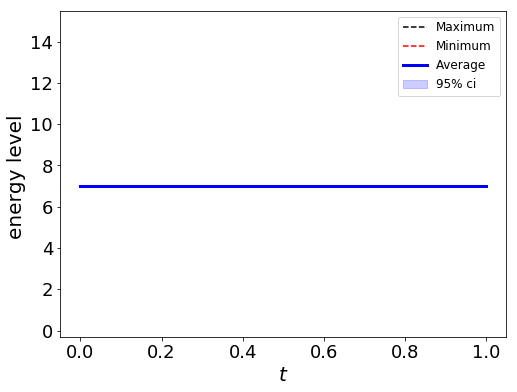} &  \includegraphics[width=5.3cm]{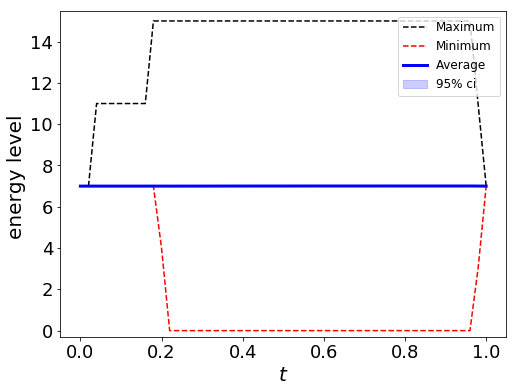} \\ \includegraphics[width=5.3cm]{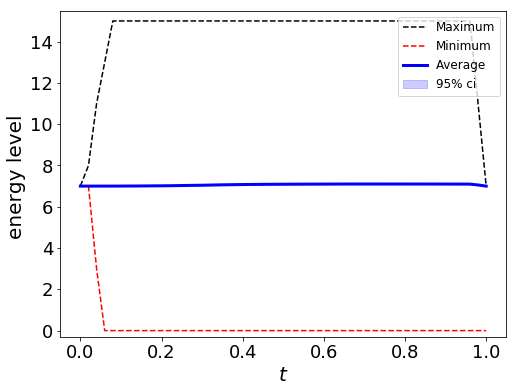} &  \includegraphics[width=5.3cm]{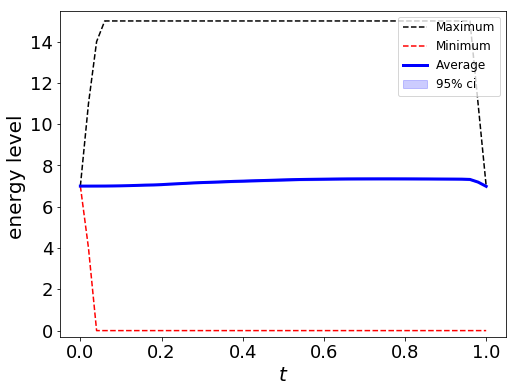}
    \end{tabular}
\caption{The average energy levels in storage for contract 1, respectively from left to right, low volatility, mid-low volatility, mid-high volatility and high volatility obtained with the LSMC method.}
\label{figure: average energy level contract 1}
\end{figure}

Python 3.7.1 is used for all the numerical experiments and the CPU is an Intel(R) Core(TM) i7-8750H CPU (2.20GHz, 2208 Mhz, 6 Cores, 12 Logical Processors).
\\
\\
\noindent \textbf{Contract 1: Standard electricity storage}\\
Table \ref{table: values contract 1} shows the numerical results of contract 1, with characteristics defined in Table \ref{table: characteristics contract 1}. In addition, in Figure \ref{figure: average energy level contract 1} the average energy levels in storage are given, together with the $95\%$ confidence intervals. 
\begin{table}[H]
\centering
\begin{tabular}{|c||c|c c c|}
\hline
Pricing method          & \multicolumn{1}{c|}{LSMC 95\% c.i.} & \multicolumn{3}{c|}{COS}                                            \\ \hhline{|=#=|===|}
$\sigma$                   & N=25000   & N=100 &  N=150   & N=200            \\ \hline
0.3                     & [0.0000 , 0.0000]&  0.0174  &  -0.0003  & 0.0000      \\
0.6                     & [-0.0005 , 0.0014] &    -0.0002  &  0.0000  &  0.0000        \\
0.9                     & [-0.0051 , 0.0222] &     0.0087 &  0.0091  &    0.0091         \\
1.2                     &  [0.1399 , 0.1943] &  0.1388    & 0.1434  &   0.1433         \\  \hline
\end{tabular}
\caption{The values of contract 1 obtained with the COS method and the LSMC method.}
\label{table: values contract 1}
\end{table}
% \begin{figure}[H]
% \centering
%     \begin{tabular}{c c c}
%         \includegraphics[width=4.5cm]{figures/col_3_ci_max_min_sigma03.png} &  \includegraphics[width=4.5cm]{figures/col_3_ci_max_min_sigma06.png} & \includegraphics[width=4.5cm]{figures/col_3_ci_max_min_sigma09.png} 
%     \end{tabular}
% \caption{The average energy level in storage for, respectively from left to right, low volatility, mid-low volatility, mid-high volatility and high volatility obtained with the LSMC method.}
% \label{figure: average energy level contract 1}
% \end{figure}
Table \ref{table: values contract 1} shows that the values obtained with the COS method converge and are always in the $95\%$ LSMC confidence interval. Furthermore, the values of contract 1 are relatively low, especially for less volatile price processes. The reason for this is that the efficiency of the electricity storage is $ 95\% $, so the discounted electricity price must increase by $(1/0.95-1)\cdot 100\%$, to make a profit. This price change happens more often when the volatility of the price is high, with the low volatility price process ($\sigma=0.3$) this did not occur, as shown in Figure \ref{figure: average energy level contract 1}.

\begin{figure}[H]
\centering
    \begin{tabular}{c c}
        \includegraphics[width=5.3cm]{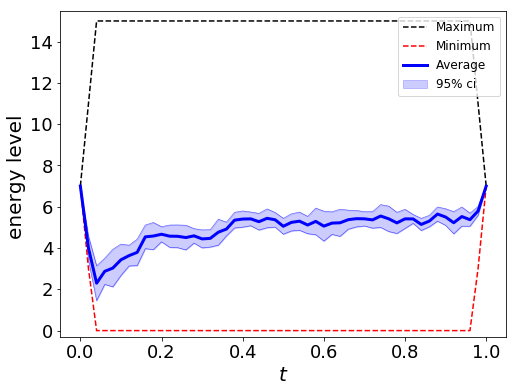} &  \includegraphics[width=5.3cm]{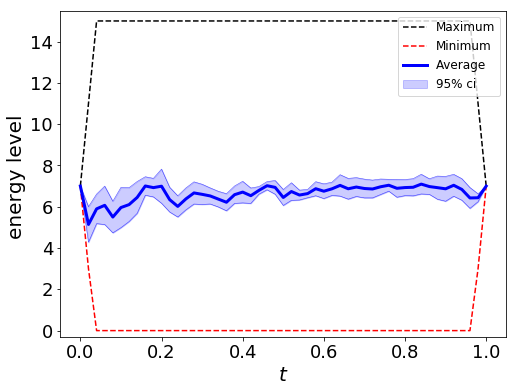} \\ \includegraphics[width=5.3cm]{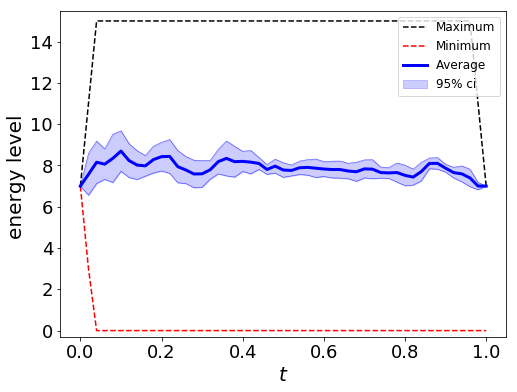} &  \includegraphics[width=5.3cm]{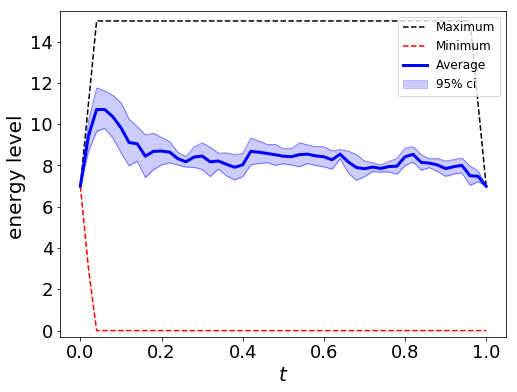}
    \end{tabular}
\caption{The average energy level in storage for contract 2, respectively from left to right, low volatility, mid-low volatility, mid-high volatility and high volatility obtained with the LSMC method.}
\label{figure: average energy level contract 2}
\end{figure}

\begin{figure}[H]
\centering
    \begin{tabular}{c c}
        \includegraphics[width=5.3cm]{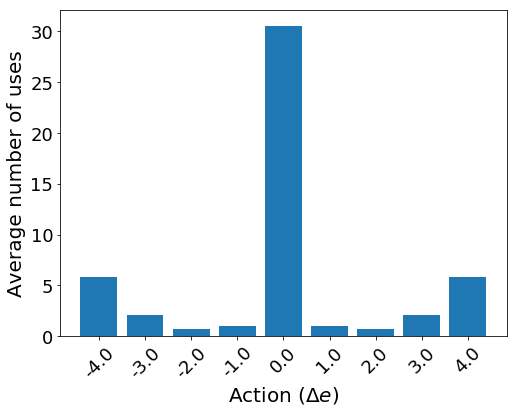} &  \includegraphics[width=5.3cm]{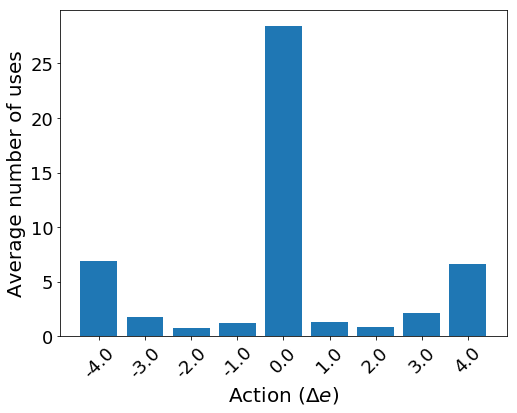} \\ \includegraphics[width=5.3cm]{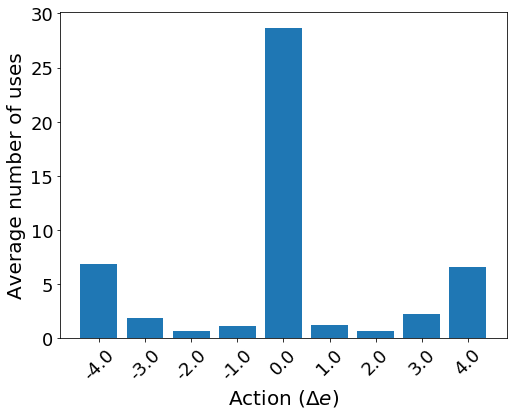} &  \includegraphics[width=5.3cm]{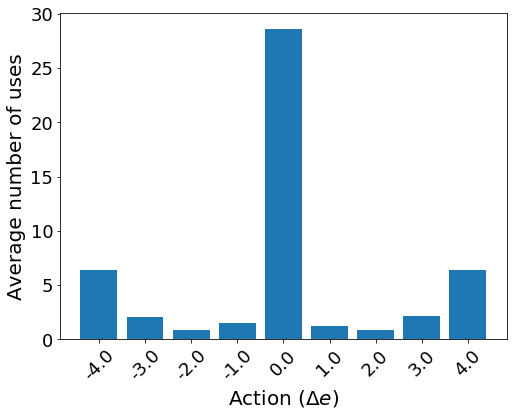}
    \end{tabular}
\caption{The average number of uses for the various actions $\Delta e \in \mathcal{A}$ obtained with the LSMC method.}
\label{figure: average uses level contract 2}
\end{figure}

Although there is little energy change taking place, a zoom of Figure \ref{figure: average energy level contract 1} shows that the strategy is to slowly start storing electricity around $t = 0.1$, if this is expected to yield a profit. In the end, the energy is usually sold until the starting energy level is reached to avoid being fined $350$ euros at the settlement date. In Figure \ref{figure: average energy level contract 1}, with mid-high and high volatility levels, all energy is sometimes sold from the battery even if a penalty has to be paid, because this strategy appears to yield more profit then.
\\
\\
\noindent  \textbf{Contract 2: Highly efficient electricity storage}\\
Table \ref{table:values contract 2} and Figures \ref{figure: average energy level contract 2} and \ref{figure: average uses level contract 2} state the numerical results for contract 2.
\begin{table}[H]
\centering
\begin{tabular}{|c||c|c c c|}
\hline
Pricing method          & \multicolumn{1}{c|}{LSMC 95\% c.i.} & \multicolumn{3}{c|}{COS}                                            \\ \hhline{|=#=|===|}
$\sigma$                   & N=25000   & N=100 &  N=150   & N=200            \\ \hline
0.3                     & [1.8550 , 1.9254] &  1.9301 & 1.8624  &  1.8630       \\
0.6                     & [3.4642 , 3.6050] &  3.4770  &  3.4640  &  3.4641        \\
0.9                     & [5.2075 , 5.4154] &   5.2304 &  5.2291  &    5.2291         \\
1.2                     &  [7.1293 , 7.3802] &  7.1323   & 7.1465  &   7.1464          \\ \hline
\end{tabular}
\caption{The values of contract 2 obtained with the COS method and LSMC method.}
\label{table:values contract 2}
\end{table}

As shown in Table \ref{table:values contract 2}, the values obtained with the COS method converge and are within the $ 95\%$ LSMC confidence intervals. The values of contract 2 are significantly higher than those of contract 1, because the $100\%$ battery efficiency allows the holder of the contract to make greater use of small fluctuations in the price process.

\begin{figure}[H]
\centering
    \begin{tabular}{c c}
        \includegraphics[width=5.3cm]{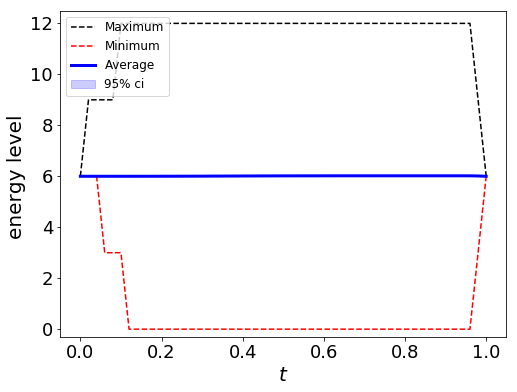} &  \includegraphics[width=5.5cm]{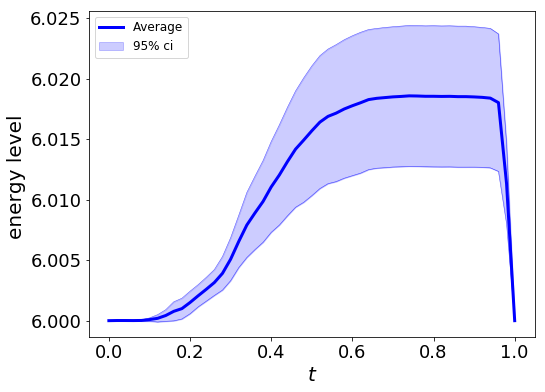} 
    \end{tabular}
\caption{The average energy level in storage for contract 3 for high volatility obtained with the LSMC method.}
\label{figure: average energy level contract 5}
\end{figure}

The strategy is different from contract 1 and depends heavily on the volatility parameter. We see a gradual shift in Figure \ref{figure: average energy level contract 2} from first releasing electricity and then loading at the low volatility parameter to loading electricity first and then releasing at the high volatility parameter. Figure \ref{figure: average uses level contract 2} shows that a very similar number of actions is taken for the different volatility parameters. As the settlement date approaches, it is ensured that the energy level returns to the starting energy level to avoid the penalty. The maximum and minimum energy levels in Figure \ref{figure: average energy level contract 2} show that at mid-high and high volatility electricity price processes all electricity is sometimes sold and the penalty is accepted.
\\
\\
\noindent \textbf{Contract 3: Car-Park as Power Plant (CPPP)}\\
In Table \ref{table:values contract 5} and Figure \ref{figure: average energy level contract 5} the numerical results are shown for contract 3, intended for the concept CPPP. 
\begin{table}[H]
\centering
\begin{tabular}{|c||c|c c c|}
\hline
Pricing method          & \multicolumn{1}{c|}{LSMC 95\% c.i.} & \multicolumn{3}{c|}{COS}                                            \\ \hhline{|=#=|===|}
$\sigma$                   & N=25000   & N=100 &  N=150   & N=200            \\ \hline
0.3                     & [0.0000 , 0.0000] &  0.0114  & -0.0002  &   0.0000    \\
0.6                     & [-0.0001 , 0.0000] &  0.0000  &  0.0000  &  0.0000        \\
0.9                     & [-0.0008 , 0.0012] &  -0.0001 &  0.0000  &   0.0000        \\
1.2                     &  [-0.0044 , 0.0020] &  0.0000   & 0.0004   &   0.0004          \\ \hline
\end{tabular}
\caption{The values of contract 3 obtained with the COS method and the LSMC method.}
\label{table:values contract 5}
\end{table}

Table \ref{table:values contract 5} shows that the CPPP has almost no profit by trading on the electricity market, where the electricity prices are generated by the dynamics described in Equations~(\ref{S_t electricity contract})-(\ref{parameters electricity contract}). The reason is that the efficiency is not high enough to make profitable use of the fluctuations of the electricity price (this can change due to technical improvements). However, an electricity storage has additional value, e.g. guaranteeing a more stable and reliable energy system.

\begin{figure}[H]
\centering
    \begin{tabular}{c c}
        \includegraphics[width=5.3cm]{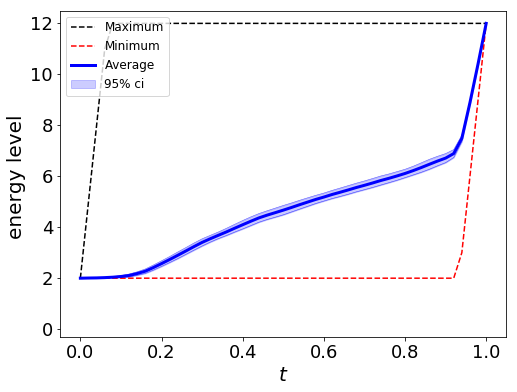} &  \includegraphics[width=5.3cm]{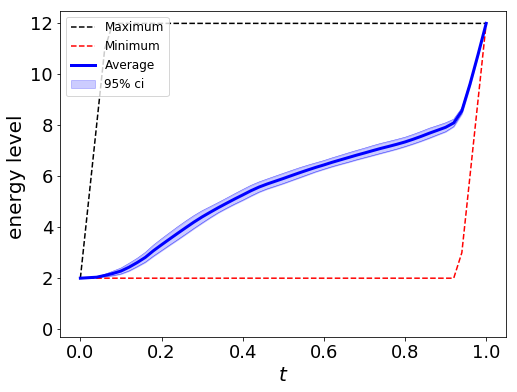} \\ \includegraphics[width=5.3cm]{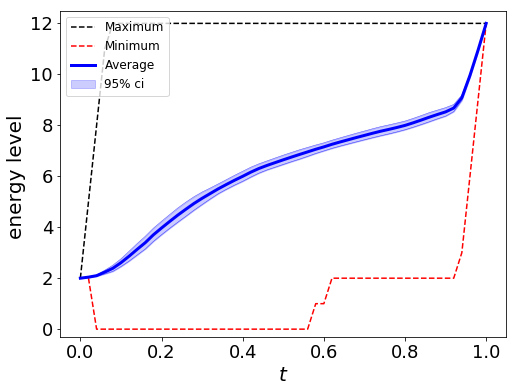} &  \includegraphics[width=5.3cm]{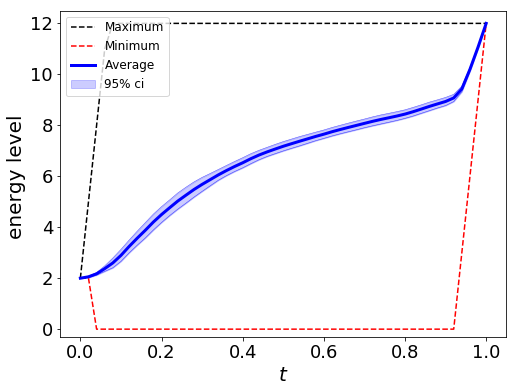}
    \end{tabular}
\caption{The average energy level in storage for contract 4, respectively from left to right, low volatility, mid-low volatility, mid-high volatility and high volatility obtained with the LSMC method.}
\label{figure: average energy level contract 4 buy sell}
\end{figure}

\begin{figure}[H]
\centering
    \begin{tabular}{c c}
        \includegraphics[width=5.3cm]{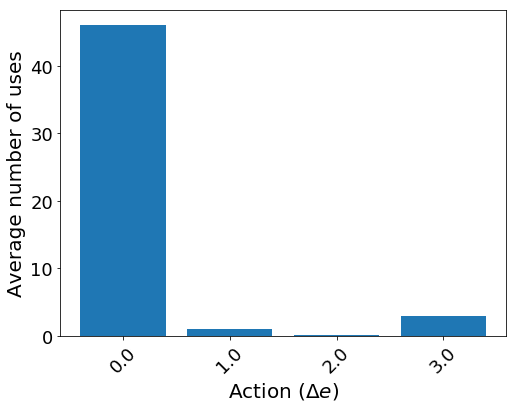} &  \includegraphics[width=5.3cm]{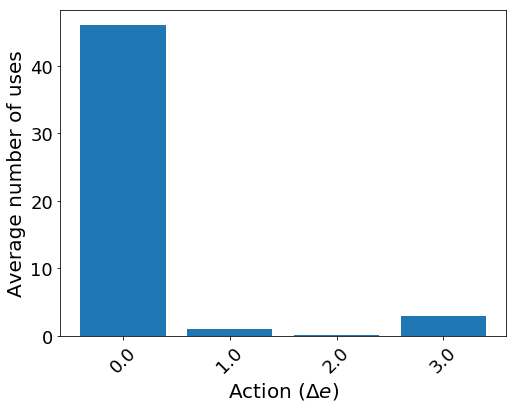} \\ \includegraphics[width=5.3cm]{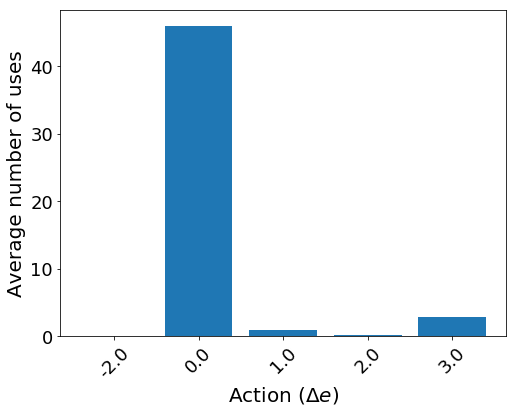} &  \includegraphics[width=5.3cm]{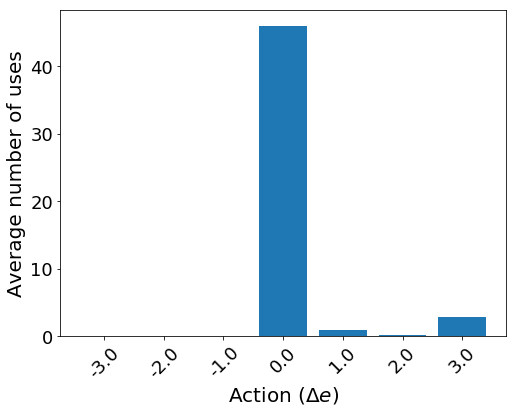}
    \end{tabular}
\caption{The average number of uses for the various actions $\Delta e \in \mathcal{A}$ obtained with the LSMC method.}
\label{figure: average energy uses contract 4 buy and sell}
\end{figure}

In Figure \ref{figure: average energy level contract 5}, the average energy level in the storage is shown, where the high volatility price process is used. The strategy is similar to that of contract 1, where the storage has an efficiency of $95\%$. A difference with the results for contract 1 is that, due to the high penalty at the settlement date, there are no trajectories in which all electricity has been sold in the end.\\
\\
\noindent \textbf{Contract 4: Cost optimization of charging electric vehicles}\\
Table \ref{table:values contract 4 buy sell} and Figures \ref{figure: average energy level contract 4 buy sell} and~\ref{figure: average energy uses contract 4 buy and sell} give the numerical results of contract 4, where the characteristics are described in Table \ref{table: characteristics contract 4:cheap charge}. 
\begin{table}[H]
\centering
\begin{tabular}{|c||c|c c c|}
\hline
Pricing method          & \multicolumn{1}{c|}{LSMC 95\% c.i.} & \multicolumn{3}{c|}{COS}                                            \\ \hhline{|=#=|===|}
$\sigma$                   & N=25000         & N=100 &  N=150   & N=200            \\ \hline
0.3                     & [-331.3365 , -331.2007]     &   -331.3426  &  -331.3153     & -331.3160 \\
0.6                     & [-330.7876 , -330.5472]   &  -330.7729    &  -330.7741  &  -330.7742         \\
0.9                     &  [-330.3961 , -330.0825] &  -329.3769  &  -330.3782 &    -330.3782          \\
1.2                     & [-330.1435 , -329.7515]    & -330.1309    & -330.1443    &    -330.1442           \\  \hline
\end{tabular}
\caption{The values of contract 4 obtained with the COS method and the LSMC method.}
\label{table:values contract 4 buy sell}
\end{table}
Table \ref{table:values contract 4 buy sell} shows that the values obtained with the COS method are always within the $95\%$ LSMC confidence intervals and are converging. The more volatile the electricity price, the cheaper the holder of the contract can charge the batteries by using the fluctuations in the price process. If we charge all cars immediately the costs will be $333.33$ euros, taking into account the physical and operational constraints, $2$-$3$ euros can be saved by the price-based charging approach if the electricity price follows the dynamics described in Equations (\ref{S_t electricity contract}) and~(\ref{parameters electricity contract}).% However, with an efficiency of $95\%$ the costs decrease to $312.76$, with $97\%$ efficiency to about $306.28$ and with $100\%$ efficiency even to $295.24$, if the high-volatile electricity price process is assumed.

The average strategy of contract 4 is to gradually buy electricity and fully charge the battery before the settlement date so that no penalty has to be paid. Moreover, with a low to mid-low volatility parameter only electricity is bought (not sold), due to the low efficiency. In contrast, with mid-high and high volatility price processes electricity is also sold, but not to a large extent as shown in Figure \ref{figure: average energy uses contract 4 buy and sell}.

%We also evaluated a contract where only charging is allowed and not the selling of electricity, while the rest of the contract characteristics remain the same. The results showed that the right to sell ensures that the costs of fully charging the battery are overall lower, even at the lower volatility parameters where no electricity is sold, while it was allowed.

\subsection{The Greeks of the electricity storage contracts}
\begin{figure}[H]
\centering
        \includegraphics[width=14.2cm]{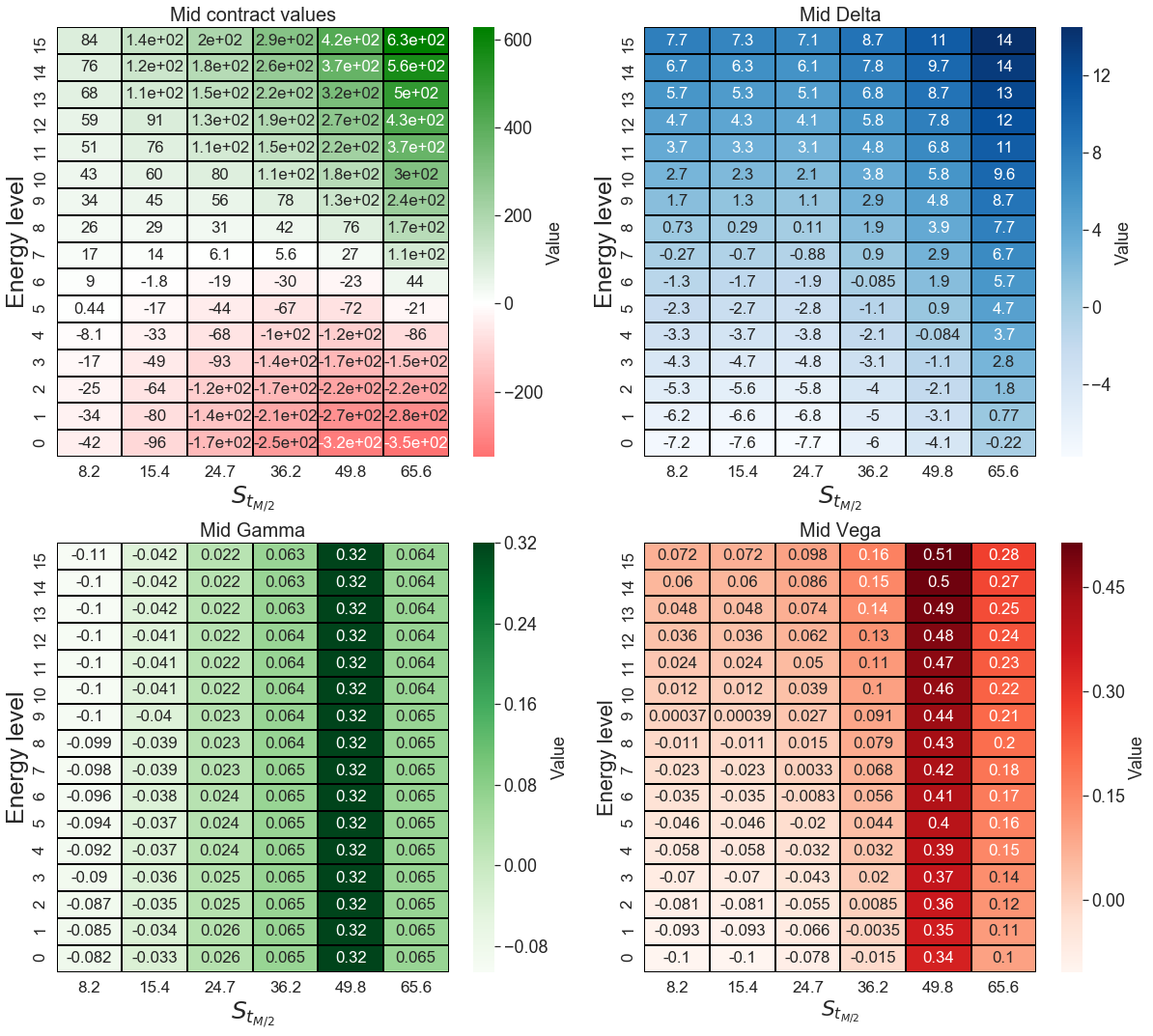} 
\caption{The Greeks of contract 2 at half the contract time $t_{M/2}$, where high volatility ($\sigma=1.2$) electricity prices are used, obtained with the COS method.}
\label{figure: mid greeks}
\end{figure}

An advantage of the COS method is that the Greeks can be calculated at almost no additional computational costs on top of the computation of the option value, unlike Monte Carlo methods. Tables \ref{table:Greeks results 03} and \ref{table:Greeks results 09} show the Greeks of the electricity contracts at initial time $t_0$, with respectively an underlying mid-low and high volatile electricity price. Moreover, in Figure \ref{figure: mid greeks} the Greeks of contract 2 are shown, as a heat map, at half the contract time, after the actions have been taken, at different fixed electricity prices with the high volatility price parameter and energy levels.
\begin{table}[H]
\centering
\begin{tabular}{c|c|c|c|c|}
\cline{2-5}
\multicolumn{1}{l|}{}       & Contract 1 & Contract 2 & Contract 3 & Contract 4 \\ \hline
\multicolumn{1}{|c|}{$\hat{\Delta}$} &    0.0000      &    0.1663       &    0.0000 &  -9.1176\\ \hline
\multicolumn{1}{|c|}{$\hat{\Gamma}$}       &    0.0001       &   0.8336         &   0.0000    & 0.4957    \\ \hline
\multicolumn{1}{|c|}{$\hat{\nu}$}    &    0.0000           &    0.3054         &    0.0000    & 0.1260      \\ \hline
\end{tabular}
\caption{The Greeks of the electricity contracts at initial time, where mid-low volatility ($\sigma=0.6$) electricity prices are used, obtained with the COS method.}
\label{table:Greeks results 03}
\end{table}

\begin{table}[H]
\centering
\begin{tabular}{c|c|c|c|c|}
\cline{2-5}
\multicolumn{1}{l|}{}       & Contract 1 & Contract 2 &  Contract 3 & Contract 4 \\ \hline
\multicolumn{1}{|c|}{$\hat{\Delta}$}  &     -0.0443       &    -0.2294         &   -0.0003   &   -9.3865  \\ \hline
\multicolumn{1}{|c|}{$\hat{\Gamma}$}    &     0.0516       &    0.4055       &  0.0003  &   0.3245     \\ \hline
\multicolumn{1}{|c|}{$\hat{\nu}$}           &     0.0372       &     0.2934         &   0.0002     &   0.1237    \\ \hline
\end{tabular}
\caption{The Greeks of the electricity contracts at initial time, where high volatility ($\sigma=1.2$) electricity prices are used, obtained with the COS method.}
\label{table:Greeks results 09}
\end{table}

Delta, $\hat{\Delta}$, measures the change in the contract value resulting from a change in the electricity price. With an electricity storage contract the energy can be bought and sold, and it may thus benefit from increasing and decreasing prices, therefore Delta can be both positive and negative. With a positive Delta, such as with contract 2 with a mid-low volatility parameter, the contract benefits from a price increase, and vice versa with a negative Delta. %Moreover, as shown in Tables \ref{table:Greeks results 03} and \ref{table:Greeks results 09}, most considered contracts benefit from a price decrease, especially highly efficient electricity storages where the price process has a high volatility parameter. %Naturally, the cost optimization of charging a vehicle benefits from a price drop, and thus $\Delta \ll 0$.

If the starting energy level is lower than the level for which the holder receives a penalty, e.g. with contract 4, the Delta drops rapidly and is negative. %This makes sense because the holder will benefit from a low price, in order not to receive the penalty at the settlement date. 
The opposite is true if the starting level is higher than the electricity level for which the penalty is obtained.

Gamma $\hat{\Gamma}$ is a measure of the change of Delta resulting from a change in the electricity price. For values of Gamma far from zero, the Delta changes drastically when the price changes, making the contract behave differently after a next price change. 

The contracts where a highly efficient storage is considered have a higher Gamma value. In addition, the Gamma also becomes slightly higher if the holder may choose actions with high energy level changes. Gamma is highest for contracts with highly efficient electricity storage and for highly volatility price processes. 

Vega $\hat{\nu}$ measures the impact of a change in the volatility parameter $\sigma$ on the contract value. The results in Tables \ref{table:Greeks results 03} and \ref{table:Greeks results 09} show that the Vega is highest for contracts that benefit from a highly volatile price process, i.e. high efficiency and allowing bigger energy changes.

Furthermore, the high values of Vega and Gamma at $S_{t_{M/2}} = 49.8$ in Figure \ref{figure: mid greeks} are because around this electricity price a choice is made to either sell all and accept the penalty or purchase a high enough energy level to avoid the penalty.

\subsection{Insights and implications of the numerical results}

The numerical results of the contract value obtained with the COS method are all within the LSMC $95\%$ confidence intervals. From this it can be implied that the COS method accurately approximates the contract values. %In addition, an advantage of the COS method is that the Greeks can be computed at almost no additional computational costs.

\begin{figure}[H]
\centering
        \includegraphics[width=7cm]{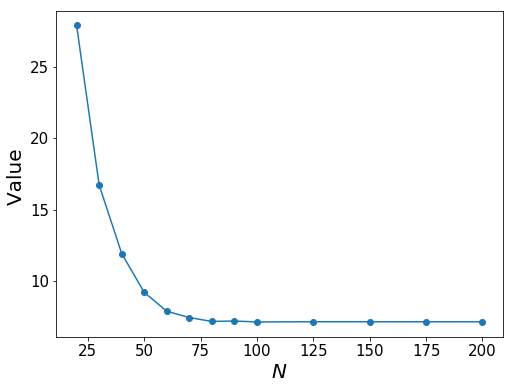} 
\caption{The value of contract 2 obtained with the COS method for different number of terms $N$ in the Fourier series expansion.}
\label{figure: convergence contract 2}
\end{figure}

A numerical check of the method's accuracy shows that the values of the considered contracts determined by the COS method converge exponentially, e.g. shown in Figure \ref{figure: convergence contract 2} for contract 2, which confirms the theoretical findings for early-exercise options in~\cite{fang2009pricing}.

Although the contract value is accurately approximated by means of the COS method, the CPU time is relatively high compared to pricing other financial derivatives, such as European and Bermudan-style options. The reason for this is that when calculating the electricity storage contract, the integral is split into many parts to determine the maximum of all actions and the continuation value must be calculated for each split, which takes most time. However, each integral can be computed in parallel.

The electricity price dynamics (\ref{S_t electricity contract}) have an impact on the electricity contract value. The larger the volatility parameter $\sigma$, the more the holder of the electricity storage contract can take advantage of large price fluctuations, resulting in a higher contract value. It is also clearly visible in the Greek $\hat{\nu}$ that the parameter $\sigma$ of the underlying price process has a major influence on the contract value, see Tables \ref{table:Greeks results 03}-\ref{table:Greeks results 09} and Figure \ref{figure: mid greeks}. In addition, the rate of mean reversion parameter $\kappa$ was examined, which shows that the contract values typically decrease as $\kappa$ increases, because for a higher $\kappa$ the price process converges faster to the mean value.

The contract value also depends on the electricity storage contract characteristics, defined in Tables \ref{table: general parameters}-\ref{table: characteristics contract 5}. The numerical results clearly show that the efficiency has a significant influence, this is because when the electricity is bought the discounted price must increase by $(1 /\eta-1)\cdot 100 \%$ in order not to make a loss. As a result, the price of the storage contract, which only takes into account trading on the electricity market by buying and selling electricity, is lower for less efficient storage. In addition, the magnitude of energy change per exercise moment is important for the value, the greater the permitted energy change, the greater the value.

Besides the dependence on the contract values, the characteristics have a major influence on the strategy of charging/selling electricity at each exercise moment. When the characteristics of the contract change, the strategy can change completely, as can be seen by comparing the strategies of contract 1 and contract 2. The price dynamics also have an effect on this. Especially the volatility parameter, not only does the strategy completely change by taking a different volatility parameter, see Figure \ref{figure: average energy level contract 2}, the confidence intervals also become smaller with a smaller volatility parameter.

\section{Conclusion} \label{sec:conclusion}
In this paper, we formalized mathematically contracts for storing electricity by trading on the electricity market, allowing the contracts to be valued efficiently. The contract takes into account the physical limitations of an electricity storage, the operational constraints of the electricity grid and the subsequent actions that a holder of the contract can make on the electricity market.

To value the electricity storage contract, we employed an efficient and robust Fourier-based pricing method, where the electricity prices follow a structural model based on a polynomial process. In particular, we focused on the well-known Ornstein-Uhlenbeck (OU) process as the underlying stochastic process. The numerical results show that the COS method performs well and can price the discussed electricity storage contract accurately. Moreover, with the COS method the Greeks can be easily computed, at any time during the contract's lifetime.

Different types of electricity storage contracts are valued, where each contract had its own characteristics (i.e. storage efficiency,  charge/discharge rate, endurance of the storage). Moreover, we looked at the concept Car Park as Power Plant (CPPP), where the numerical results indicate that the efficiency of electric vehicle batteries strongly impacts the profitability. Furthermore, the contract can be used to reduce the costs of charging electricity storage or electric vehicles. With these contracts, it appears that the efficiency of electricity storage and the volatility of the electricity price are of great influence on whether it is useful to apply electricity storage to make a profit by trading on the electricity market. 
%Also for optimizing the costs to charge an electric vehicle, the efficiency is of great importance, an efficiency of the car battery of $90\%$ and $97\%$ has a total cost saving of $1\%$ and $10\%$ respectively.
 
%An advantage of the COS method is that the Greeks can be computed at almost no additional computational costs, in contrast to the LSMC method where the calculation of the Greeks can be difficult and time consuming. However, if a general trading strategy based on the average energy level in storage is preferred over the Greeks, then the LSMC method should be used next to the COS method.

\bibliographystyle{abbrv}
\bibliography{bib/MacroStrings,bib/Articles,bib/Books} 

\begin{thebibliography}{10}

\bibitem{ahmadian2018plug}
A.~Ahmadian, M.~Sedghi, A.~Elkamel, M.~Fowler, and M.~A. Golkar.
\newblock Plug-in electric vehicle batteries degradation modeling for smart
  grid studies: Review, assessment and conceptual framework.
\newblock {\em Renewable and Sustainable Energy Reviews}, 81:2609--2624, 2018.

\bibitem{barlow2002diffusion}
M.~Barlow.
\newblock A diffusion model for electricity prices.
\newblock {\em Mathematical finance}, 12(4):287--298, 2002.

\bibitem{benth2008stochastic}
F.~Benth, J.~Benth, and S.~Koekebakker.
\newblock {\em Stochastic modelling of electricity and related markets},
  volume~11.
\newblock World Scientific, 2008.

\bibitem{boogert2008gas}
A.~Boogert and C.~De~Jong.
\newblock Gas storage valuation using a monte carlo method.
\newblock {\em The journal of derivatives}, 15(3):81--98, 2008.

\bibitem{carmona2014survey}
R.~Carmona and M.~Coulon.
\newblock A survey of commodity markets and structural models for electricity
  prices.
\newblock In {\em Quantitative Energy Finance}, pages 41--83. Springer, 2014.

\bibitem{cartea2005pricing}
A.~Cartea and M.~Figueroa.
\newblock Pricing in electricity markets: a mean reverting jump diffusion model
  with seasonality.
\newblock {\em Applied Mathematical Finance}, 12(4):313--335, 2005.

\bibitem{chen2009progress}
H.~Chen, T.~Cong, W.~Yang, C.~Tan, Y.~Li, and Y.~Ding.
\newblock Progress in electrical energy storage system: A critical review.
\newblock {\em Progress in natural science}, 19(3):291--312, 2009.

\bibitem{connolly2011practical}
D.~Connolly, H.~Lund, P.~Finn, B.~Mathiesen, and M.~Leahy.
\newblock Practical operation strategies for pumped hydroelectric energy
  storage (phes) utilising electricity price arbitrage.
\newblock {\em Energy Policy}, 39(7):4189--4196, 2011.

\bibitem{doob1942brownian}
J.~Doob.
\newblock The brownian movement and stochastic equations.
\newblock {\em Annals of Mathematics}, pages 351--369, 1942.

\bibitem{dorr2003valuation}
U.~D{\"o}rr.
\newblock Valuation of swing options and examination of exercise strategies by
  monte carlo techniques.
\newblock {\em Mathematical Finance}, 10:27, 2003.

\bibitem{duffie2000transform}
D.~Duffie, J.~Pan, and K.~Singleton.
\newblock Transform analysis and asset pricing for affine jump-diffusions.
\newblock {\em Econometrica}, 68(6):1343--1376, 2000.

\bibitem{dunn2011electrical}
B.~Dunn, H.~Kamath, and J.-M. Tarascon.
\newblock Electrical energy storage for the grid: a battery of choices.
\newblock {\em Science}, 334(6058):928--935, 2011.

\bibitem{evans2012assessment}
A.~Evans, V.~Strezov, and T.~J. Evans.
\newblock Assessment of utility energy storage options for increased renewable
  energy penetration.
\newblock {\em Renewable and Sustainable Energy Reviews}, 16(6):4141--4147,
  2012.

\bibitem{eydeland2003energy}
A.~Eydeland and K.~Wolyniec.
\newblock {\em Energy and power risk management: New developments in modeling,
  pricing, and hedging}, volume 206.
\newblock John Wiley \& Sons, 2003.

\bibitem{fang2009novel}
F.~Fang and C.~Oosterlee.
\newblock A novel pricing method for european options based on fourier-cosine
  series expansions.
\newblock {\em SIAM Journal on Scientific Computing}, 31(2):826--848, 2009.

\bibitem{fang2009pricing}
F.~Fang and C.~Oosterlee.
\newblock Pricing early-exercise and discrete barrier options by fourier-cosine
  series expansions.
\newblock {\em Numerische Mathematik}, 114(1):27, 2009.

\bibitem{filipovic2016polynomial}
D.~Filipovi{\'c} and M.~Larsson.
\newblock Polynomial diffusions and applications in finance.
\newblock {\em Finance and Stochastics}, 20(4):931--972, 2016.

\bibitem{filipovic2019polynomial}
D.~Filipovi{\'c} and M.~Larsson.
\newblock Polynomial jump-diffusion models.
\newblock {\em Swiss Finance Institute Research Paper}, (17-60), 2019.

\bibitem{Warefilipovic2018polynomial}
D.~Filipovi{\'c}, M.~Larsson, and T.~Ware.
\newblock Polynomial processes for power prices.
\newblock {\em Swiss Finance Institute Research Paper}, (18-34), 2018.

\bibitem{girish2013determinants}
G.~Girish and S.~Vijayalakshmi.
\newblock Determinants of electricity price in competitive power market.
\newblock {\em International Journal of Business and Management}, 8(21):70,
  2013.

\bibitem{graves1999opportunities}
F.~Graves, T.~Jenkin, and D.~Murphy.
\newblock Opportunities for electricity storage in deregulating markets.
\newblock {\em The Electricity Journal}, 12(8):46--56, 1999.

\bibitem{julch2016comparison}
V.~J{\"u}lch.
\newblock Comparison of electricity storage options using levelized cost of
  storage (lcos) method.
\newblock {\em Applied Energy}, 183:1594--1606, 2016.

\bibitem{karlin1949geometry}
S.~Karlin and L.~Shapley.
\newblock Geometry of reduced moment spaces.
\newblock {\em Proceedings of the National Academy of Sciences of the United
  States of America}, 35(12):673, 1949.

\bibitem{kempton1997electric}
W.~Kempton and S.~Letendre.
\newblock Electric vehicles as a new power source for electric utilities.
\newblock {\em Transportation Research Part D: Transport and Environment},
  2(3):157--175, 1997.

\bibitem{kempton2001vehicle}
W.~Kempton, J.~Tomic, S.~Letendre, A.~Brooks, and T.~Lipman.
\newblock Vehicle-to-grid power: battery, hybrid, and fuel cell vehicles as
  resources for distributed electric power in california.
\newblock 2001.

\bibitem{koller2013defining}
M.~Koller, T.~Borsche, A.~Ulbig, and G.~Andersson.
\newblock Defining a degradation cost function for optimal control of a battery
  energy storage system.
\newblock In {\em 2013 IEEE Grenoble Conference}, pages 1--6. IEEE, 2013.

\bibitem{kyriakopoulos2016electrical}
G.~Kyriakopoulos and G.~Arabatzis.
\newblock Electrical energy storage systems in electricity generation: Energy
  policies, innovative technologies, and regulatory regimes.
\newblock {\em Renewable and Sustainable Energy Reviews}, 56:1044--1067, 2016.

\bibitem{lechtenbohmer2015decarbonising}
S.~Lechtenb{\"o}hmer, L.~Nilsson, M.~{\AA}hman, and C.~Schneider.
\newblock Decarbonising the energy intensive basic materials industry through
  electrification--implications for future eu electricity demand.
\newblock {\em Dubrovnik}, 2015.

\bibitem{longstaff2001valuing}
F.~Longstaff and E.~Schwartz.
\newblock Valuing american options by simulation: a simple least-squares
  approach.
\newblock {\em The review of financial studies}, 14(1):113--147, 2001.

\bibitem{lucia2002electricity}
J.~Lucia and E.~Schwartz.
\newblock Electricity prices and power derivatives: Evidence from the nordic
  power exchange.
\newblock {\em Review of derivatives research}, 5(1):5--50, 2002.

\bibitem{lund2008integration}
H.~Lund and W.~Kempton.
\newblock Integration of renewable energy into the transport and electricity
  sectors through v2g.
\newblock {\em Energy policy}, 36(9):3578--3587, 2008.

\bibitem{lund2009role}
H.~Lund and G.~Salgi.
\newblock The role of compressed air energy storage (caes) in future
  sustainable energy systems.
\newblock {\em Energy conversion and management}, 50(5):1172--1179, 2009.

\bibitem{lunz2012influence}
B.~Lunz, Z.~Yan, J.~Gerschler, and D.~Sauer.
\newblock Influence of plug-in hybrid electric vehicle charging strategies on
  charging and battery degradation costs.
\newblock {\em Energy Policy}, 46:511--519, 2012.

\bibitem{luo2015overview}
X.~Luo, J.~Wang, M.~Dooner, and J.~Clarke.
\newblock Overview of current development in electrical energy storage
  technologies and the application potential in power system operation.
\newblock {\em Applied energy}, 137:511--536, 2015.

\bibitem{OosterleeGrzelak201911}
C.~Oosterlee and L.~Grzelak.
\newblock {\em {Mathematical Modeling and Computation in Finance}}.
\newblock World Scientific, first edition, November 2019.
\newblock {ISBN 978-1-78634-794-7}.

\bibitem{palizban2016energy}
O.~Palizban and K.~Kauhaniemi.
\newblock Energy storage systems in modern grids—matrix of technologies and
  applications.
\newblock {\em Journal of Energy Storage}, 6:248--259, 2016.

\bibitem{ParkLee}
E.~Park~Lee.
\newblock {\em A socio-technical exploration of the Car as Power Plant}.
\newblock PhD thesis, Delft University of Technology, 2019.

\bibitem{poullikkas2013comparative}
A.~Poullikkas.
\newblock A comparative overview of large-scale battery systems for electricity
  storage.
\newblock {\em Renewable and Sustainable Energy Reviews}, 27:778--788, 2013.

\bibitem{schwartz2000short}
E.~Schwartz and J.~Smith.
\newblock Short-term variations and long-term dynamics in commodity prices.
\newblock {\em Management Science}, 46(7):893--911, 2000.

\bibitem{staffell2016maximising}
I.~Staffell and M.~Rustomji.
\newblock Maximising the value of electricity storage.
\newblock {\em Journal of Energy Storage}, 8:212--225, 2016.

\bibitem{sun2019polynomial}
Z.~Sun.
\newblock Polynomial processes for energy markets.
\newblock 2019.
\newblock PowerPoint presentation.

\bibitem{von2015benchop}
L.~von Sydow, L.~Josef~H{\"o}{\"o}k, E.~Larsson, E.~Lindstr{\"o}m,
  S.~Milovanovi{\'c}, J.~Persson, V.~Shcherbakov, Y.~Shpolyanskiy,
  S.~Sir{\'e}n, J.~Toivanen, et~al.
\newblock Benchop--the benchmarking project in option pricing.
\newblock {\em International Journal of Computer Mathematics},
  92(12):2361--2379, 2015.

\bibitem{weron2014electricity}
R.~Weron.
\newblock Electricity price forecasting: A review of the state-of-the-art with
  a look into the future.
\newblock {\em International journal of forecasting}, 30(4):1030--1081, 2014.

\bibitem{zhang2012efficient}
B.~Zhang, L.~Grzelak, and C.~Oosterlee.
\newblock Efficient pricing of commodity options with early-exercise under the
  ornstein--uhlenbeck process.
\newblock {\em Applied Numerical Mathematics}, 62(2):91--111, 2012.

\bibitem{zhang2013efficient}
B.~Zhang and C.~Oosterlee.
\newblock An efficient pricing algorithm for swing options based on fourier
  cosine expansions.
\newblock {\em Journal of Computational Finance}, 16(4):1--32, 2013.

\end{thebibliography}

 \newpage
 \appendix
 \section{Increasing polynomial map}\label{appendix:increasing polynomial map}
 To model spot prices with the polynomial model a polynomial map is needed that provides a 1-1 map from a polynomial process $X_t$ to the spot price $S_t$. In order to capture the characteristics of the electricity commodity market, increasing polynomial maps $\Phi(\ \cdot \ )$ from $[0,\infty)$ onto $[0,\infty)$ will be used.

Such increasing polynomial maps can be constructed from non-negative polynomials on $[0,\infty)$ \cite{karlin1949geometry}. To create these increasing polynomial maps the multiplication of positive quadratics will be used. These quadratics are defined as \cite{sun2019polynomial}:
\begin{equation}
    \tilde{q}_{\alpha,\gamma} (x) = \frac{\alpha}{2} x^2 + (1-\alpha - \gamma)x + \gamma,
    \label{non-neg-polynomial}
\end{equation}
where the quadratic polynomials $\tilde{q}_{\alpha,\gamma}$ are normalized such that:
$$\int_0^\infty e^{-x}\tilde{q}_{\alpha,\gamma}(x)dx = 1.$$
To produce all the possible quadratic polynomials that are on $[0,\infty)$, the parameter set $(\alpha, \gamma)$ is chosen as follows:
\begin{equation*}
        \alpha = \hat{r} \cos (\xi),
\end{equation*}
\begin{equation*}
        \gamma = \hat{r} \sin(\xi),
\end{equation*}
where $\xi \in \left[0,\pi / 2 \right] $ and $\hat{r}\in \left[0, \cos(\xi) + \sin(\xi) +\sqrt{\sin(2\xi)}\right] $.
\\
\\
With the quadratic polynomials as described above, the increasing polynomial map, $\Phi(\ \cdot \ ) : [0,\infty) \longmapsto [0,\infty)$, can be constructed from pairs of parameters $(\alpha_1,\gamma_1),...,(\alpha_K,\gamma_K)$ where $K\in \mathbb{N}^+$:
\begin{equation}
    \Phi(x)=\int_{0}^x \prod_{k=1}^K \tilde{q}_{\alpha_k,\gamma_k} (u) du.
    \label{polynomial_map}
\end{equation}
Note that if $\alpha_k \neq 0$ $\forall k\in [1,K]$ the degree of $\Phi$ is $2K+1$. Furthermore, under this general construction of the polynomial map (\ref{polynomial_map}) a polynomial process with even order can be obtained by setting $\alpha_k=0$ for some $k \in [1,K]$.
 \section{Ornstein-Uhlenbeck process}\label{appendix:OU process}
 The Ornstein-Uhlenbeck (OU) process has the following SDE:
\begin{equation}
    dX_t=\kappa (\theta-X_t)dt +\sigma dW_t,
    \label{OU_process_dynamics}
\end{equation}
where $t\geq 0$, $\kappa$ the rate of mean reversion, $\theta$ the long-run mean, $\sigma$ the volatility of the process and $W_t$ the Brownian motion under real-world measure $\mathbb{P}$.

Integration from $t_0$ to $t$ on both sides of the SDE (\ref{OU_process_dynamics}) yields the following solution of the OU process:
\begin{equation}
    X_t=X_{t_0}e^{-\kappa(t-t_0)}+\theta\left(1-e^{-\kappa (t-t_0)}\right)+\sigma e^{-\kappa (t-t_0)}\int_{t_0}^t e^{\kappa s} dW_s.
    \label{solution X_t OU process appendix}
\end{equation}
By using a scaled time-transformed Brownian motion an analytic solution of the integral in equation (\ref{solution X_t OU process appendix}) can be computed, see \cite{doob1942brownian} for further details. 

The OU process is normally distributed, i.e. $X_t\sim N\big(\mathbb{E}[X_t],Var[X_t]\big)$, where the expected value and variance are given by:
\begin{align}
    \begin{split}
\mathbb{E}[X_t|\mathcal{F}_0]&=X_{t_0} e^{-\kappa (t-t_0)}+\theta (1- e^{-\kappa (t-t_0)}),\\
Var[X_t|\mathcal{F}_0]&=\frac{\sigma^2}{2\kappa}(1-e^{-2\kappa (t-t_0)}),
    \end{split}
    \label{literature_exp_variance_OU}
\end{align}
with initial position $X_{t_0}$.

The well-known characteristic function of the OU process is given by \cite{zhang2012efficient}:
\begin{equation}
    \phi(u|x,\Delta t)=e^{iuxe^{-\kappa \Delta t}+ A(u,\Delta t)},
\end{equation}
where 
$$A(u,\Delta t)=\frac{1}{4\kappa}\left(e^{-2\kappa \Delta t}-e^{-\kappa \Delta t}\right)\left(u^2\sigma^2+u e^{\kappa \Delta t}\left(u \sigma^2-4i\kappa \theta\right)\right).$$

\section{LSMC algorithm}\label{appendix: LSMC algorithm}
In this appendix, we briefly explain the Least Squares Monte Carlo (LSMC) algorithm to price the electricity storage contract. This LSMC method presents a simple, intuitive, yet powerful approximation to value early-exercise options, e.g. Bermudan options, swing options and electricity storage contracts. The main insight behind LSMC is that the conditional expectation to compute the continuation value can be approximated by the use of least squares regression as a function of the simulated price process (see \cite{longstaff2001valuing} and \cite{dorr2003valuation}). An advantage of LSMC is that it is not model dependent, a multi-factor pricing model can be valued just as easily as a simpler model.    

The following notation is used for the algorithm:
\begin{itemize}
\itemsep0.25em 
    \item $S^{i}_m$ : The asset price of trajectory $i$ at time $t_m$.
    \item $CV_m^{i,e}$: The continuation value with $e\in E$ energy level in storage after action $\Delta e \in \mathcal{A}$ is taken at time $t_m$.
    \item $CF_m^{i,e}$: The cash flow at time $t_m$ with $e\in E$ energy level in storage.
    \item $ACF_m^{i,e}$: The accumulated value of the cash flows for the optimal actions realised for moments $t_k$, $k\in\{m,...,M+1\}$, with energy level $e\in E$.
    \item $DACF_m^{i,e}:= e^{-r\Delta t}ACF_{m+1}^{i,e}$, the discounted accumulated value of cash flows with $e\in E$ energy in storage.
    \item $Q^{\Delta e}_m  := q_b(\Delta e)$, the penalty function, which is imposed if an action $\Delta e\in \mathcal{A}(t_m,e)\setminus\mathcal{D}(t_m,e)$ is taken.
    \item $PO^{i,\Delta e}_m:=g(t_m,S^i_m,\Delta e)$, the payoff function if action $\Delta e\in\mathcal{A}$ is taken, as defined in formula (\ref{payoff function energy contract}).
\end{itemize}

Moreover, two functions of the Python package NumPy are used for the regression, polyfit and polyval. The polyfit function applies a polynomial regression which minimizes the error in a least squares sense. After the regression coefficients are obtained with the polyfit function, we can use polyval to evaluate the polynomial function on other data points to get a prediction.

\begin{algorithm}[H]
\caption{The LSMC algorithm for the electricity contract.}
\begin{algorithmic}[1]
\For {$e=0,...,N_{cap}$}
    \State $CF_{M+1}^{i,e}=q_s(t_{M+1},S^i(t_{M+1}),e)$ \ \ \ \ ,$\forall i=1,...N$
    \State $ACF_{M+1}^{i,e}=CF_{M+1}^{i,e}$  \ \ \ \ ,$\forall i=1,...N$
\EndFor
\For {$m=M,...,1$}
    \State $X^i=S^i(t_m)$ \ \ \ \ ,$\forall i=1,...,N$
    \For {$e=0,...,N_{cap}$}
        \State $DACF^{i,e}=e^{-r\Delta t}ACF_{m+1}^{i,e}$ \ \ \ \ ,$\forall i=1,...,N$
        \State $p^e=\text{polyfit}(X,DACF^{e},3)$
        \State $CV^{i,e}_m=\text{polyval}(p^e,X^i_m)$ \ \ \ \ ,$\forall i=1,...,N$
    \EndFor
    \For {$e=0,...,N_{cap}$}
        \For {$i=1,...,N$}
            \State $\Delta e^{i,*}=\argmax_{\Delta e \in \mathcal{A}(t_m,e)}\big\{PO_m^{i,\Delta e} + CV_m^{i,e+\Delta e} + Q^{\Delta e} \big\}$ 
            \State $CF^{i,e}_{m}=PO_m^{i,\Delta e^{i,*}} + Q^{\Delta e^{i,*}}$ 
            \State $ACF^{i,e}_{m}=CF^{i,e}_{m} + e^{-r\Delta t} ACF^{i, e+\Delta e^{i,*}}_{m+1}$ 
        \EndFor
    \EndFor
\EndFor
\For  {$e=0,...,N_{cap}$}
\State $v(t_0,S(t_0), e) = \frac{1}{N}\sum_{i=1}^N DACF_0^{i,e}= \frac{1}{N}\sum_{i=1}^N e^{-r\Delta t}ACF_1^{i,e}$
\EndFor
\end{algorithmic}
\label{Algorithm electricity contract}
\end{algorithm}

\end{document}